\documentclass[12pt]{iopart}

\usepackage{amsmath}
\usepackage{amssymb}
\usepackage{graphicx}
\usepackage{float}
\usepackage{caption}
\usepackage{wrapfig}
\usepackage{adjustbox}
\usepackage{subfigure}
\usepackage{booktabs}
\usepackage{array}
\usepackage{colortbl}
\usepackage{csquotes}
\usepackage{changepage}
\usepackage{wrapfig}
\usepackage{cite}

\begin{document}

%\title[JOREK fast particle tracker]{Simulating runaway electrons in realistic magnetohydrodynamic fields: introduction of the fast particle in JOREK code}

%\title{Simulation of relativistic test electron transport in a JET disruption}

\title{Electron acceleration in a JET disruption simulation}

\author{C. Sommariva\textsuperscript{1}\footnote{Present address: \'Ecole Polytechnique F\'ed\'erale de Lausanne (EPFL), Swiss Plasma Center (SPC), CH-1015, Lausanne, Switzerland}, E. Nardon\textsuperscript{1}, P. Beyer\textsuperscript{2}, M. Hoelzl\textsuperscript{3}, G. T. A. Huijsmans\textsuperscript{1,4} and JET Contributors\textsuperscript{5,*}}

\address{\textsuperscript{1}CEA, IRFM, F-13108, Saint Paul-lez-Durance, France}
\address{\textsuperscript{2}Aix-Marseille Universit\'e, CNRS, PIIM UMR 7345, 13397, Marseille Cedex 20, France}
\address{\textsuperscript{3}Max Planck Institute for Plasma Physics, Boltzmannstr. 2, 85748 Garching b. M., Germany}
\address{\textsuperscript{4}Dep. Applied Physics, T. U. Eindhoven, P. O. B. 513, 5600, Eindhoven, Netherlands}
\address{\textsuperscript{5}EUROfusion Consortium, JET, Culham Science Center, Abingdon, OX14 3DB, UK}
\address{\textsuperscript{*}See the author list of  “X. Litaudon et al 2017 Nucl. Fusion 57 102001}
\ead{cristian.sommariva@epfl.ch}
\vspace{10pt}
\begin{indented}
\item[]January 2018
\vspace{2pc}
\submitto{\NF}
\end{indented}
\maketitle

\begin{abstract}
Runaways are suprathermal electrons having sufficiently high energy to be continuously accelerated up to tens of MeV by a driving electric field \cite{connor75}. Highly energetic runaway electron (RE) beams capable of damaging the tokamak first wall can be observed after a plasma disruption \cite{reux15}. Therefore, it is of primary importance to fully understand their generation mechanisms in order to design mitigation systems able to guarantee safe tokamak operations. In a previous work, \cite{sommariva18}, a test particle tracker was introduced in the JOREK 3D non-linear MHD code and used for studying the electron confinement during a simulated JET-like disruption. It was found in \cite{sommariva18} that relativistic electrons are not completely deconfined by the stochastic magnetic field taking place during the disruption thermal quench (TQ). This is due to the reformation of closed magnetic surfaces at the beginning of the current quench (CQ). This result was obtained neglecting the inductive electric field in order to avoid the unrealistic particle acceleration which otherwise would have happened due to the absence of collision effects. The present paper extends \cite{sommariva18} analysing test electron dynamics in the same simulated JET-like disruption using the complete electric field. For doing so, a simplified collision model is introduced in the particle tracker guiding center equations. We show that electrons at thermal energies can become RE during or promptly after the TQ due to a combination of three phenomena: a first REs acceleration during the TQ due to the presence of a complex MHD-induced electric field, particle reconfinement caused by the fast reformation of closed magnetic surfaces after the TQ and a secondary acceleration induced by the CQ electric field.
%
 %are able to survive the disruption thermal quench (TQ) chaotic magnetic field thanks to the reformation of closed magnetic surfaces at the current quench (CQ) beginning. This result was obtained neglecting the inductive electric field in order to avoid the unrealistic particle acceleration which otherwise would have happened due to the absence of collision effects. The present paper extends \cite{sommariva18} analysing test electron dynamics in the same simulated JET-like disruption using the complete electric field. For doing so, a simplified collision model is introduced in the particle tracker guiding center equations. We show that electrons at thermal energies can become RE during or promptly after the TQ due to a combination of three phenomena: a first REs acceleration during the TQ due to the presence of a complex MHD-induced electric field, particle reconfinement caused by the fast reformation of closed magnetic surfaces after the TQ and a secondary acceleration induced by the CQ electric field.
\end{abstract}

%\ioptwocol

\section{Introduction} \label{chp4_electron_acceleration}

In tokamak plasmas, fast electrons are said to run away when collision effects are not capable to compensate the electron acceleration induced by a driving electric field \cite{dreicer59}. Highly energetic runaway electron (RE) beams are sometimes experimentally observed during plasma disruptions with possible harmful consequences for the reactor first wall \cite{reux15}. Indeed, a plasma facing component (PFC) struck by RE might suffer damages due to the deposition of high heat loads \cite{reux15}. The energy carried by these suprathermal electrons can considerably increase with the plasma current ($\mathrm{I_p}$). Therefore, their presence during high $\mathrm{I_p}$ discharges in ITER ($\mathrm{I_p \approx 15 MA}$) has to be considered as a serious threat to PFCs \cite{sugihara12,lehnen14,hollmann15}. For this reason, RE prevention and mitigation systems for the ITER tokamak are under advanced state of design \cite{sugihara12,lehnen14}. Ideally, their development should be based on the complete understanding of the physics underlying the formation and dissipation of REs. Unfortunately, complete answers to questions concerning the mechanisms and plasma configurations allowing the generation of a disruptive initial (primary) RE seed have not been achieved yet. 

Generally, two consequent phases characterise a tokamak disruption: the first one is called thermal quench (TQ) and consists of a fast (few milliseconds) and almost complete loss of the plasma thermal energy. The second one, known as current quench (CQ), is identified by a decrease of $\mathrm{I_p}$ which leads to discharge termination. This $\mathrm{I_p}$ reduction is imputable to the very large post-TQ plasma resistance \cite{hender07} and induces a toroidal driving electric field. At the end of the CQ, disruptive runaway beams may be observed via a slowly decaying $\mathrm{I_p}$ which correlates with different radiation measurements such as: synchrotron, soft and hard X-rays, gamma and neutron emissions \cite{reux15,abdullaev16}. The disruption runaway phase, which is not systematically seen in experiments, is known as runaway plateau \cite{reux15,abdullaev16} and can last up to few hundreds of milliseconds.

The generation mechanism of disruptive REs can be decomposed into two different families: the first one is the primary generation which consists of all the processes capable to produce REs without requiring an already existing relativistic electron population \cite{martin-solis17}. On the other hand, the second one, called the secondary generation or electron avalanche \cite{rosenbluth97,nilsson15}, exponentially amplifies an already existing primary RE seed via knock-on collisions between thermal and relativistic charge carriers \cite{rosenbluth97,nilsson15}. In ITER, the electron avalanche is thought to be one of the dominant generation processes but, unfortunately, estimations of the maximum runaway beam current are difficult to obtain. One of the reasons of such a high task complexity is the significant secondary generation sensitivity to the primary RE current and by large incertitudes on the prediction of these last. This is justified by the ITER large avalanche amplification factor which exponentially magnify the errors coming from RE seed current estimations to RE beam scales. As an example of the expected exponential growth, a $\mathrm{\sim 2\cdot 10^{-8}MA}$ seed is foreseen to produce RE beams having currents up to $\mathrm{\sim2MA}$ by electron avalanche in a ITER 15MA disruption being characterised by a CQ time of $\mathrm{\sim 50ms}$ and mitigated by argon mixed with $\mathrm{7kPa\cdot m^3}$ of deuterium \cite{martin-solis17}. These considerations highlight the importance to better understand the processes underlying the primary RE generation in order to achieve a more effective disruption mitigation system design.

At the moment, four different mechanisms are identified as belonging to the primary generation: Dreicer, Hot Tail, tritium $\beta$-decay and the Compton scattering \cite{martin-solis17}. The Dreicer mechanism \cite{dreicer59,connor75}, which is strictly related to the subject of this paper, consists in the acceleration of thermal electrons up to relativistic energies due to the presence of a driving electric force stronger than the average Coulomb collision drag. In disruptions, this electric field is generally associated to the one caused by the CQ $\mathrm{I_p}$ decay \cite{martin-solis17} but it may also be induced by the virulent MHD activity taking place during the TQ. The CQ Dreicer generation is foreseen not to significantly contribute to the formation of REs in ITER \cite{martin-solis17} while the TQ one was not taken into account in previous work. Differently, the Hot Tail mechanism involves the acceleration of a supra-thermal electron distributions emerging from the TQ phase which is thought to be caused by the incomplete thermalisation of the pre-TQ ones \cite{smith05,smith08}. The Hot Tail is foreseen to be one of the main primary generation mechanisms in ITER mitigated disruptions \cite{martin-solis17}. The last two processes, which are respectively the hot electron emission due to tritium $\beta$-decay and the thermal electron acceleration caused by collisions with energetic photons emitted by activated wall components, have not been experimentally observed yet but they are thought to contribute to the generation of primary REs in ITER \cite{martin-solis17}.   

% up to runaway conditions (Dreicer generation mechanism \cite{dreicer59,connor75}). 
%
%In ITER, the CQ electric field is foreseen to be small enough to prevent the Dreicer generation \cite{martin-solis17}. This does not imply that ITER plasma disruptions will be RE free. Indeed, the electron avalanche phenomenon, which is the generation of fast electrons due to knock-on collisions between thermal and runaway ones \cite{rosenbluth97,nilsson15}, is capable to create a high current beam exponentially amplifying an already existing suprathermal electron population \cite{solis152}. The primary RE seed required for onsetting the avalanche can be obtained from a variety of generation processes, e.g. the tritium $\beta$-decay and the Compton scattering between electrons and energetic photons emitted by activated wall components \cite{martin-solis17}. Of primary importance is the mechanism called Hot Tail \cite{smith05,smith08} which consists in the production of suprathermal electron populations after the TQ due to the incomplete thermalisation of the pre-TQ ones. Indeed, \cite{martin-solis17} estimates a 10MA RE current due to Hot Tail generation as worst scenarios for a ITER 15MA disruption.

As discussed above, in ITER the $\mathrm{I_p}$ decay induced electric field is foreseen to be small enough to prevent the CQ Dreicer generation \cite{martin-solis17} but little is known about the possibility of obtaining RE due to the strong MHD activity taking place during the TQ. This possibility requires, at least, the satisfaction of two conditions: the presence of an electric field strong enough to drive electrons up to high energies and a sufficiently long electron confinement time. The second condition was already addressed theoretically in \cite{boozer161,boozer162} and numerically in \cite{sommariva18}. Indeed, in \cite{boozer161,boozer162} it is shown that if the plasma magnetic topology during the TQ presents respectively residual closed flux surfaces at the edge or flux tubes non intercepting the first wall, electrons can be confined long enough to become RE. On the other hand, in \cite{sommariva18} a numerical study of the electron confinement properties in a JET-like disruption simulation is presented. The latter concerns JET pulse 86887 which is an $\mathrm{I_p=2MA}$ - $\mathrm{B_0=2T}$ Ohmic discharge where a disruption was obtained via $\mathrm{D_2}$ massive gas injection (MGI). In this simulation, the MGI destabilises an MHD `modal cascade' \cite{nardon17} from a large 2/1 mode up to the complete magnetic field stochastisation obtained setting $\mathrm{q_0 > 1}$. In addition to the full magnetic field chaoticity, the `artificial' $\mathrm{q_0 > 1}$ setting has also the effect to suppress the internal kink mode observed in \cite{nardon17}. It was found in \cite{sommariva18} that electrons are able to `survive' the TQ for a wide range of initial energies and radial positions. All these elements sustain the possibility that (fast) particles are not totally deconfined by the strong MHD activity of the TQ. A pioneering study of the REs dynamics in tokamak disruption is presented in \cite{izzo11} where the confinement properties of electrons having initial energy above the RE threshold are analysed via combined NIMROD MHD-particle tracking simulations. The results presented in \cite{izzo11} also confirm the incomplete fast electron deconfinement during the TQ indeed, the RE confinement time is found to increase with $\mathrm{\sim R^3}$ (where R is the tokamak major radius). Despite the presence of both accelerating and decelerating terms in the NIMROD particle model, investigations on the electron dynamics and, more precisely, on possible RE acceleration mechanisms during a disruption TQ have not been reported yet. From the experimental side, losses of RE within energies from 1MeV to 3MeV  were observed before the CQ phase in DIII-D killer pellet-induced disruptions \cite{james12}. \cite{james12} relates these `prompt' losses to the RE generation due to the existence of a high loop voltage in the TQ probably induced by a significant increase of hyper-resistivity. Moreover, \cite{james12} shows that the spatial deposition of RE losses seen during these experiments is consistent with NIMROD disruption simulations.

For these reasons, the present work explores the electron dynamics during the TQ phase of the JOREK simulated JET pulse 86887 disruption via the fast particle tracker already introduced and exploited in \cite{sommariva18}. In contrast with \cite{sommariva18}, we focus our investigation on assessing the possibility for thermal electrons to become RE due to the presence of acceleration mechanisms all along the disruption TQ phase.

In order to do so, a guiding center (GC) collision drag force is introduced into the GC model presented in the Section 3 of \cite{sommariva18}. This JOREK fast particle tracker development is described in Section \ref{chp4_sec_collisional_drag}.  The evolution of the parallel effective electric field (sum of the parallel electric force and collision drag) during the simulated disruption is studied in Section \ref{chp4_sec_electric_field_disruption}. In Section \ref{chp4_electron_acceleration_sec_electron_accelerarion_TQ} an analysis of the particle behaviour during the TQ is given. In this phase, RE formation is observed due to electron acceleration by large local MHD-induced parallel electric fields. A study on scenarios and parameters of the MHD disruption simulations altering the particle acceleration processes is furnished in Section \ref{chp4_plasma_resistivity_and_RE}. Conclusions are presented at the end of this paper (Section \ref{chp4_summary}).
\section{A collision drag model for the JOREK Guiding Center tracker} \label{chp4_sec_collisional_drag}

The representation of the electric field used in the JOREK code is the following \cite{morales15}:

\begin{equation}
\mathbf{E} = -\nabla \Phi - \frac{\partial \psi}{\partial t}\frac{\mathbf{e}_{\phi}}{R} \label{chp4_eq_Efield}
\end{equation}

where $\mathrm{\Phi=R_0B_0u}$ is the electric potential as a function of the stream function u, the magnetic axis major radius $\mathrm{R_0}$ and the reference toroidal magnetic field $\mathrm{B_0}$, $\psi$ is the poloidal magnetic flux, $\mathbf{e}_{\phi}$ is the unit vector in the geometrical toroidal direction and R is the major radius. In a previous work \cite{sommariva18}, the inductive term ($\frac{\partial \psi}{\partial t}$) of Eq.\ref{chp4_eq_Efield} was neglected in order to avoid an unrealistic electron acceleration during the pre-TQ phase of a disruption. Indeed, the GC model used in \cite{sommariva18} did not take into account the energy dissipation of test electrons due to their collisions with the background plasma, dissipation which tends to counteract the $\frac{\partial \psi}{\partial t}$-induced acceleration. In the present work, a simplified collision drag model is introduced in the GC equations presented in \cite{sommariva18} which are reported below for sake of completeness:

\numparts
\begin{eqnarray}
\eqalign{\mathbf{\dot{X}} = \mathrm{{\frac{1}{{\mathbf{b}} \cdot {\mathbf{B^{\ast}}}}} \left({q{\mathbf{E}}{\times}{\mathbf{b}}} - {{p_{\parallel}}{\frac{{\partial}{\mathbf{b}}}{\partial t}} \times {\mathbf{b}}} + {\frac{m{\mu}{\mathbf{b}}{\times}{\mathbf{{\nabla}B}} + {p_{\parallel}}{\mathbf{B^{\ast}}}}{m \gamma_{GC}}}\right)}} \label{gc_position}  \\
\eqalign{\mathrm{\dot{p_{\parallel}}} =\mathrm{ {\frac{\mathbf{B^{\ast}}}{\mathbf{b} \cdot \mathbf{B^{\ast}}}} \cdot \left(q{\mathbf{E}}-{{p_{\parallel}}{\frac{{\partial}{\mathbf{b}}}{\partial t}}}-{\frac{{\mu}{\nabla}{B}}{\gamma_{GC}}}\right)}} \label{gc_momentum} \\
\mathrm{\gamma_{GC}} = \mathrm{\sqrt{1+{\left(\frac{p_{\parallel}}{mc}\right)^2} + {\frac{2{\mu}B}{mc^2}}}} \label{gamma_GC}
\end{eqnarray}
\endnumparts

where ${\mathbf{X}}$ is the GC position vector, $p_{\parallel}$ is the GC momentum parallel to the magnetic field, $\mathrm{{\mu}=\frac{{\| \mathbf{p}-{p_{\parallel}}\mathbf{b} \|}^2}{2mB}}$ is the magnetic moment \cite{tao07}, $\mathrm{B}$ is the magnetic field intensity, ${\mathbf{b}}={\frac{\mathbf{B}}{\mathrm{B}}}$ is the magnetic field direction, $\mathrm{{\mathbf{B^{\ast}}}={p_{\parallel}}{{\mathbf{\nabla}}\times {\mathbf{b}}}+q{\mathbf{B}}}$ is the so-called ``effective magnetic field'', $\mathrm{q}$ and $\mathrm{m}$ are respectively the particle charge and mass while $\mathrm{c}$ is the speed of light. 

The drag force used in this work is the one given in \cite{solis15} which is adapted to the JOREK MHD model including molecular deuterium ($\mathrm{D_2}$) \cite{sommariva17}:

\numparts
\begin{gather}
	\mathbf{F} = -\mathrm{\frac{q^4}{4\pi \epsilon^2_0 E_0} \frac{\gamma((\gamma + 1)\alpha_e + \alpha_i)}{(\gamma^2 - 1)^{\frac{3}{2}}}} \frac{\mathbf{p}}{\mathrm{mc}} \label{eq_chp4_drag_force_fussmann} \\
	\gamma = \sqrt{1+\left(\frac{\mathbf{p}}{\mathrm{mc}}\right)^2} \\
	\mathrm{\alpha_e = n{\;\ln{\left({\Lambda_{ef}}\right)}} + n_{D_2}{\;\ln{\left({\Lambda_{eb}}\right)}} \label{chp4_eq_electron_collisional_density}} \\
	\mathrm{\alpha_i = n{\;\ln{\left({\Lambda_{if}}\right)}} + n_{D_2}{{\left({Z_{nucl,D_2}}\right)}^2}{\;\ln{\left({\Lambda_{nucl,D_2}}\right)}} \label{chp4_eq_ion_collisional_density}} \\
	\mathrm{\Lambda_{ef} = \frac{(\gamma-1)\sqrt{\gamma+1}\lambda_D}{2\gamma r_e},} \quad \mathrm{\Lambda_{eb} = (\gamma-1)\sqrt{\gamma+1} \frac{E_0}{I_z}} \\
	\mathrm{\Lambda_{if} = \frac{(\gamma^2 - 1)\lambda_D}{\gamma r_e},} \quad \mathrm{\Lambda_{nucl,k} = \frac{(\gamma^2 - 1)E_0}{\gamma I_z}} \\
\end{gather}
\endnumparts

where $\mathbf{p}$ is the particle momentum in 3D momentum space, ${\epsilon_0}$ is the vacuum permittivity, $\mathrm{E_0=mc^2}$ is the electron rest energy, $\mathrm{n=n_{ef}=n_{if}}$ and $\mathrm{n_{D_2}}$ are respectively the background plasma and molecular deuterium impurity number densities, $\mathrm{Z_{nucl,D_2}}$ is the deuterium nuclear charge which is set equal to 2 under the assumption of simultaneous collision with the two $\mathrm{D_2}$ nuclei, $\lambda_D$ is the Debye length ($\mathrm{\lambda_D = \sqrt{\frac{\epsilon_0 k_B T_e T_i}{q^2n(T_e + T_i)}}}$), $\mathrm{r_e= q^2/(4\pi\epsilon_0E_0)}$ is the classical electron radius \cite{feynman65} and $\mathrm{I_z=15.5eV}$ is the $\mathrm{D_2}$ ionisation energy taken from \cite{linstrom17}. It has to be remarked that in Eq.\ref{chp4_eq_electron_collisional_density} and \ref{chp4_eq_ion_collisional_density} the quasi-neutrality assumption $\mathrm{n = n_i = n_e}$ of the JOREK MHD model is used (where $\mathrm{n_i}$ and $\mathrm{n_e}$ are respectively the plasma ion and electron densities).

The introduction of the drag force (Eq.\ref{eq_chp4_drag_force_fussmann}) in the JOREK fast particle tracker GC model is obtained neglecting its component acting on the particle perpendicular velocity (the magnetic moment remains an adiabatic invariant of motion) and the drag-induced drifts appearing in Eq.\ref{gc_position} resulting from the GC expansion. Moreover, the plasma fields are approximated substituting the GC position to the particle one. Thus, the modified GC parallel momentum equation is:

\numparts
\begin{gather}
\mathrm{\frac{dp_{\parallel}}{dt} = \frac{\mathbf{B^{\ast}}}{{\mathbf{B^{\ast}}}\cdot{\mathbf{b}}} \cdot \left({q{\mathbf{E}} - {p_{\parallel}}\frac{\partial {\mathbf{b}}}{\partial t} -\frac{\mu \nabla B}{\gamma_{GC}}}\right) + F_{{\parallel},coll} \label{chp4_eq_parallel_momentum_drag}} \\
\mathrm{F_{{\parallel},coll} = - {\frac{q^4}{4{\pi}{\epsilon^2_0}E_{0}}} {\frac{\gamma_{GC}{\left({{\left({\gamma_{GC} + 1}\right)}\alpha_e + \alpha_i}\right)}}{{\left({\gamma_{GC} - 1}\right)}^{\frac{3}{2}}}} \frac{p_{\parallel}}{{m}c} \label{chp4_eq_collision_drag}}
\end{gather}
\endnumparts

where $\mathrm{\alpha_e}$ and $\mathrm{\alpha_i}$ are defined by Eq.\ref{chp4_eq_electron_collisional_density} and \ref{chp4_eq_ion_collisional_density}. This simplistic model is justified by the scope of the present work. Indeed, our main aim is to perform a first assessment of the running away possibility of a thermal electron due to the TQ electric field acceleration and not to precisely describe its phase space dynamics. For this reason, we preferred the faster and more intuitive drag model composed by Eq.\ref{chp4_eq_parallel_momentum_drag} and \ref{chp4_eq_collision_drag} leaving as future work the implementation of a more accurate Monte Carlo solver such as the one presented in \cite{sarkimaki17}.

One of the drawbacks of this simple collision drag model is the absence of a `thermal bath'. We minimise its consequences initialising particle populations just before the TQ. It has also to be remarked that neither the effects of the background plasma velocity nor the ones due to plasma fluxes are taken into account by this collision drag. Indeed, the plasma rotation was found to be significantly smaller than the electron thermal velocity in the considered discharge thus, its effects on the collision drag are neglected in the present work.  

%These last can be neglected in the present work because they are found to be dominated by the `thermal bath' ones in the treated disruption simulation.

%Despite its simplicity, this minimal model is not exampted from drawbacks. Indeed, one of its shortcomes is the impossibility to maintain the particle momentum at thermal values which, instead, decreases towards zero. We minimise the consequences of this effect initialising particle populations just before the TQ.

\section{Parallel effective electric field} \label{chp4_sec_electric_field_disruption}

In order to assess whether electrons are accelerated or decelerated, the critical quantity to be considered is the net parallel force ($\mathrm{F_{\parallel}}$) acting on each particle. In the GC model used in this work, $\mathrm{F_{\parallel}}$ is given by the right hand side of Eq.\ref{chp4_eq_parallel_momentum_drag}:

\begin{equation}
	 \mathrm{F_{\parallel} = \frac{\mathbf{B^{\ast}}}{{\mathbf{B^{\ast}}}\cdot{\mathbf{b}}} \cdot \left({q{\mathbf{E}} - {p_{\parallel}}\frac{\partial {\mathbf{b}}}{\partial t} -\frac{\mu \nabla B}{\gamma_{GC}}}\right) + F_{{\parallel},coll} \label{sec3_eq_parallel_force}}
\end{equation}

For the kinetic energy levels investigated hereafter, the $\frac{\mathbf{B^{\ast}}}{{\mathbf{B^{\ast}}}\cdot{\mathbf{b}}} \cdot {p_{\parallel}}\frac{\partial {\mathbf{b}}}{\partial t}$ and $\frac{\mathbf{B^{\ast}}}{{\mathbf{B^{\ast}}}\cdot{\mathbf{b}}} \cdot \frac{\mu \nabla B}{\gamma_{GC}}$ terms have small effects on the electron distribution spreading in energy space during the TQ. Consequently, we focus our attention on the net force resulting from the parallel electric field and the collision drag:

\begin{equation}
	 \mathrm{F_{\parallel,acc} = \left(\frac{\mathbf{B^{\ast}}}{\mathbf{B^{\ast}}\cdot \mathbf{b}} \cdot q \mathbf{E} + F_{\parallel,coll}\right)} \label{sec3_eq_acc_parallel_force}
\end{equation}  

It is worth remarking that $\mathrm{F_{\parallel,acc}}$ can be interpreted in terms of a parallel effective electric field ($\mathrm{E_{\parallel,eff}}$) defined as follows:

\begin{equation}
\mathrm{E_{\parallel,eff} =\frac{1}{|q|}\left(\frac{\mathbf{B^{\ast}}}{\mathbf{B^{\ast}}\cdot \mathbf{b}} \cdot q \mathbf{E} + F_{\parallel,coll}\right) = \frac{F_{\parallel,acc}}{|q|}} \label{chp4_effective_electric_field}
\end{equation}

This quantity is particularly convenient because it allows to rewrite the collisional GC parallel momentum equation (Eq.\ref{chp4_eq_parallel_momentum_drag}) into the same form of the non-collisional one (Eq.\ref{gc_momentum}):

\begin{equation}
\mathrm{\frac{dp_{\parallel}}{dt} = |q|E_{\parallel,eff} - \frac{\mathbf{B^{\ast}}}{{\mathbf{B^{\ast}}}\cdot{\mathbf{b}}} \cdot \left({ {p_{\parallel}}\frac{\partial {\mathbf{b}}}{\partial t} +\frac{\mu \nabla B}{\gamma_{GC}}}\right)}
\end{equation}

Therefore, $\mathrm{E_{\parallel,eff}}$ can be interpreted as the electric field accelerating an `equivalent' non-collisional particle. This interpretation makes $\mathrm{E_{\parallel,eff}}$ the most suitable quantity for evaluating the capability of a MHD field to accelerate or decelerate electrons thus, it will be extensively used in the remaining of this work.   

In Tables \ref{tab:chp4_Eff_phi45E1to100keVpitch170_tearingtoTQ} the $\mathrm{E_{\parallel,eff}}$ for energies and pitch angle respectively of $\mathrm{[1,10,100]keV}$ and $170^{\circ}$ are reported at different times in the disruption simulation. This pitch angle value is chosen within the typical experimental interval of $\theta \approx \left(5^{\circ},12^{\circ}\right)$ seen in various machines \cite{jaspers01,zychen06,stahl13} and respects the RE counter-current motion \cite{izzo11,papp111} (the JET plasma current and magnetic field are both in clockwise direction seen from above thus, electron counter current motions are characterised by a negative $\mathrm{p_{\parallel}}$). It has also to be remarked that this setting allows comparisons between the analysis reported in the following of this paper and the one of \cite{sommariva18}.

\begin{table}[H]
	%	\centering
	\begin{adjustwidth}{-1.9cm}{}
		\begin{tabular}{|>{\centering\arraybackslash}m{3.77cm}|>{\centering\arraybackslash}m{3.77cm}|>{\centering\arraybackslash}m{3.77cm}|>{\centering\arraybackslash}m{4.2cm}|>{\centering\arraybackslash}m{3.0cm}|}
			\toprule
			$\mathrm{E_{kin}=1keV}$ & $\mathrm{E_{kin}=10keV}$ & $\mathrm{E_{kin}=100keV}$ & Poincar\'e plot & time and phase \\
			\midrule
			\includegraphics[width=3.77cm, height=4.75cm]{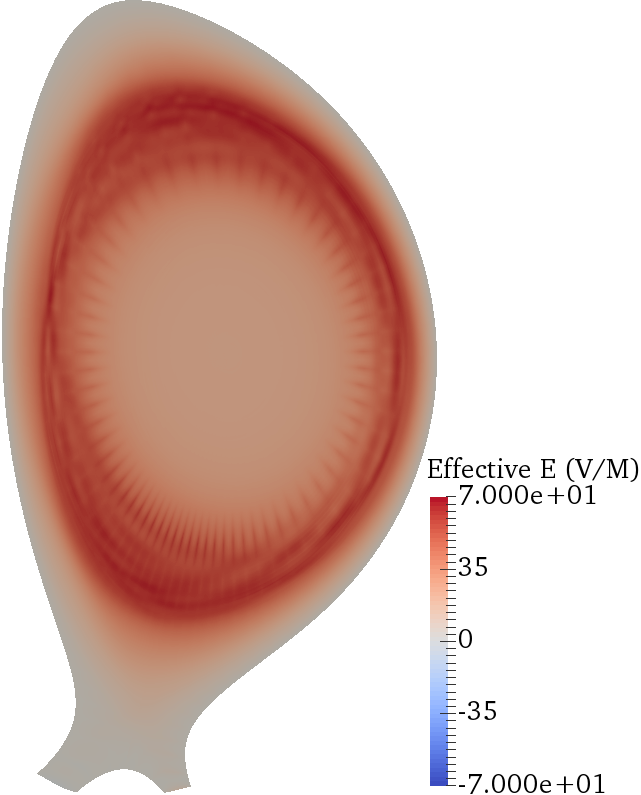} & \includegraphics[width=3.77cm, height=4.75cm]{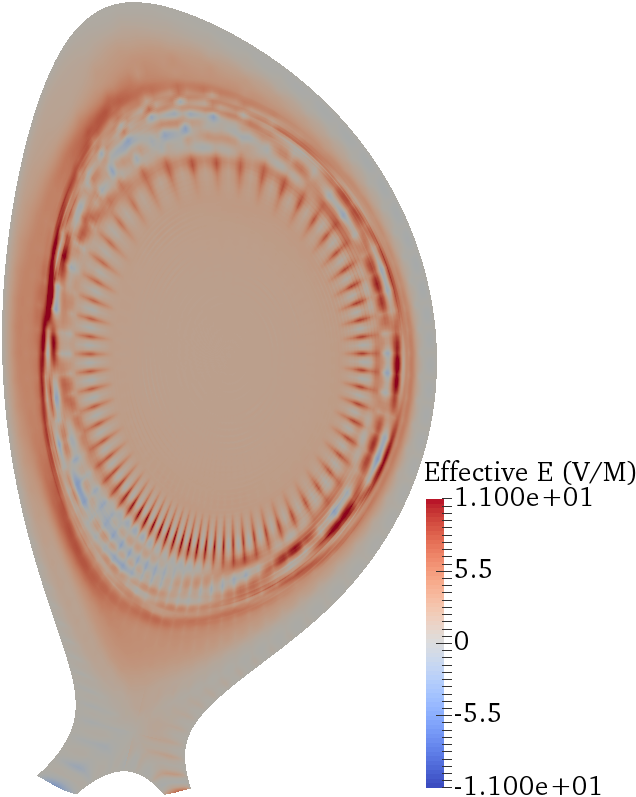} & \includegraphics[width=3.77cm, height=4.75cm]{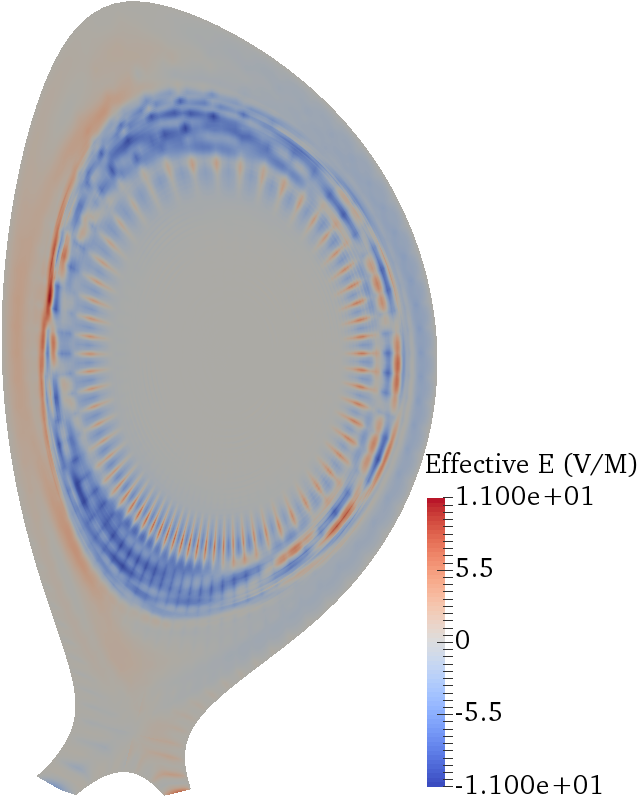} & \includegraphics[width=3.5cm, height=4.75cm]{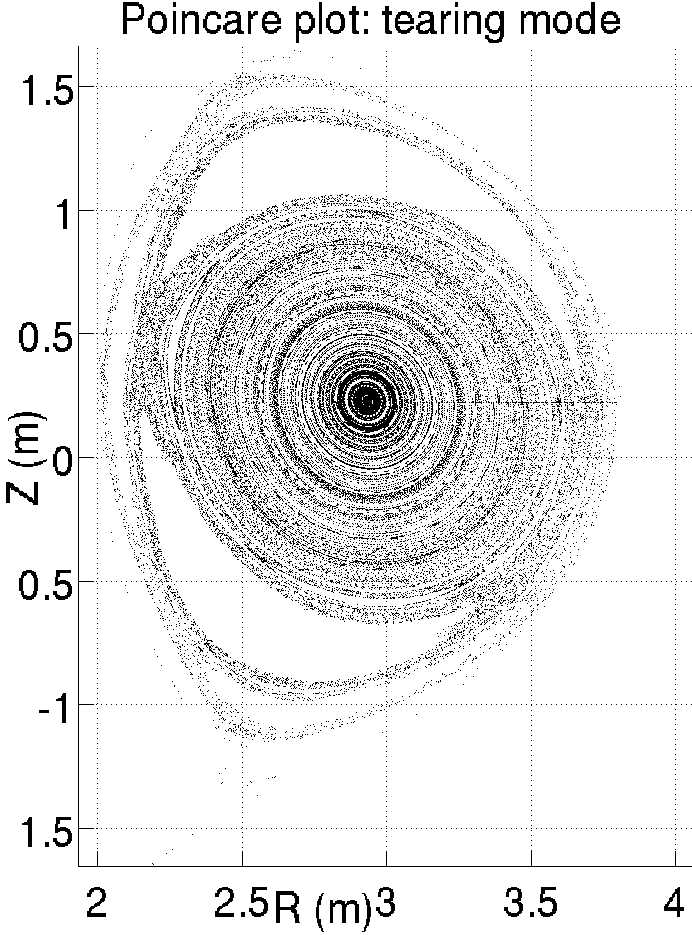} & {\begin{tabular}{c} t=3.55ms: \\ pre-TQ with \\ large 2/1 \\ magnetic island \end{tabular}} \\
			\midrule
			\includegraphics[width=3.77cm, height=4.75cm]{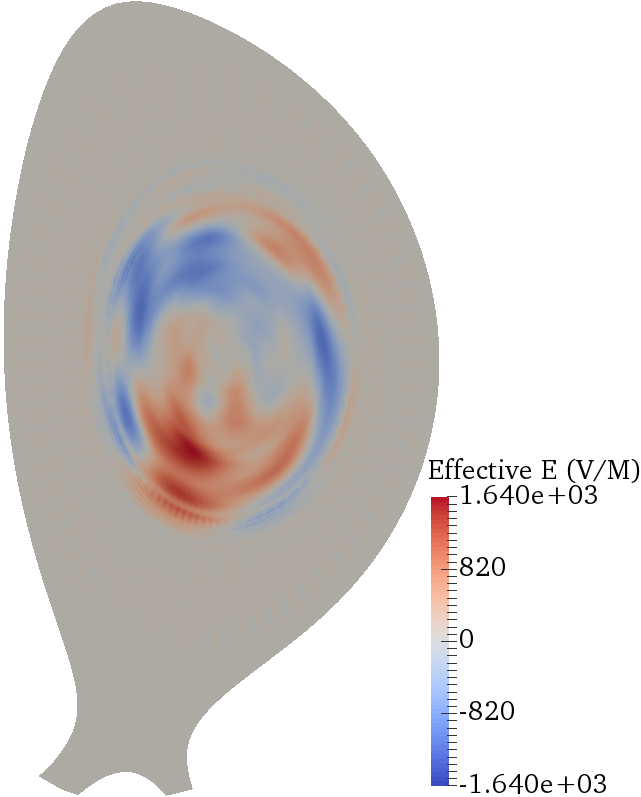} & \includegraphics[width=3.77cm, height=4.75cm]{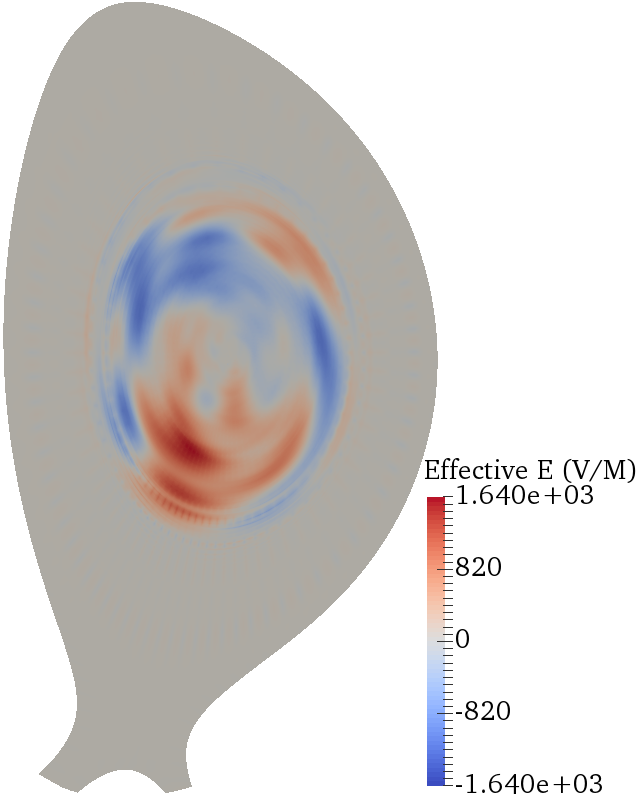} & \includegraphics[width=3.77cm, height=4.75cm]{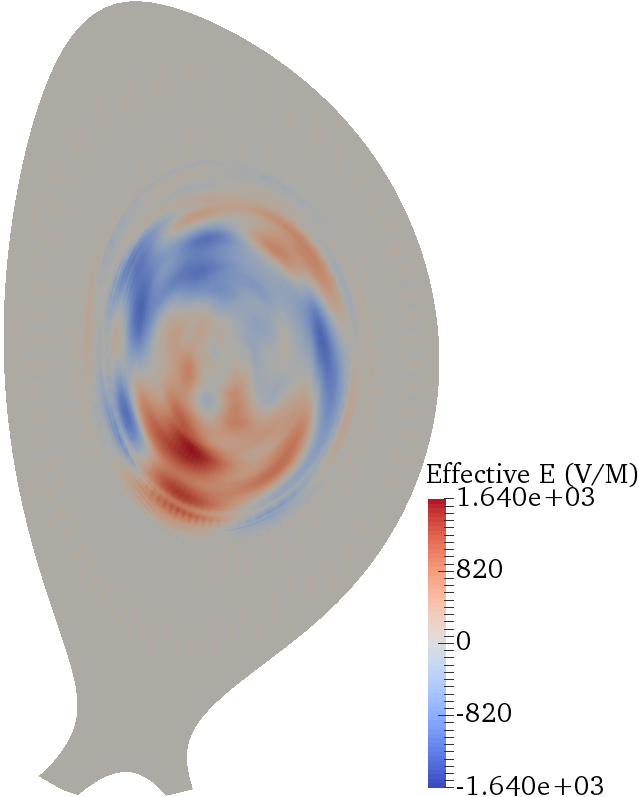} & \includegraphics[width=4.25cm, height=4.75cm]{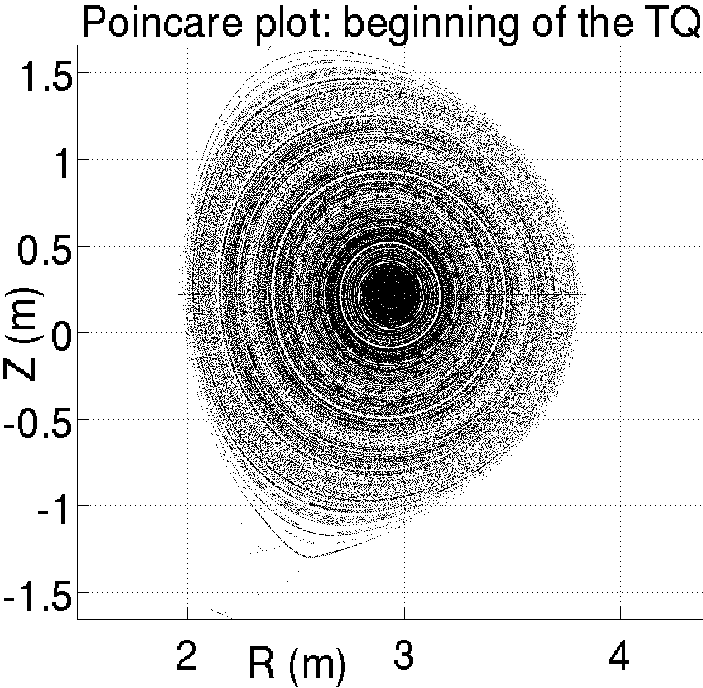} & {\begin{tabular}{c} t=3.83ms: \\ TQ beginning \end{tabular}} \\
			\midrule
			\includegraphics[width=3.77cm, height=4.75cm]{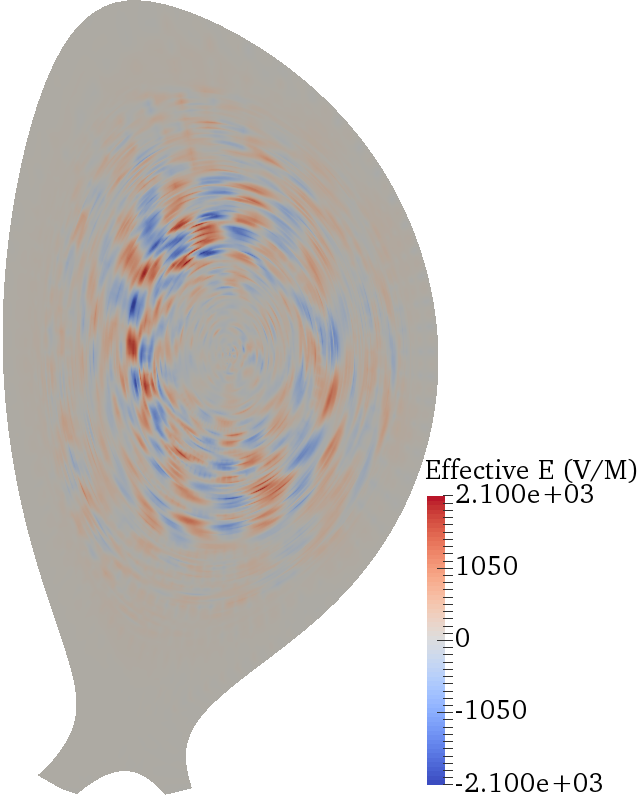} & \includegraphics[width=3.77cm, height=4.75cm]{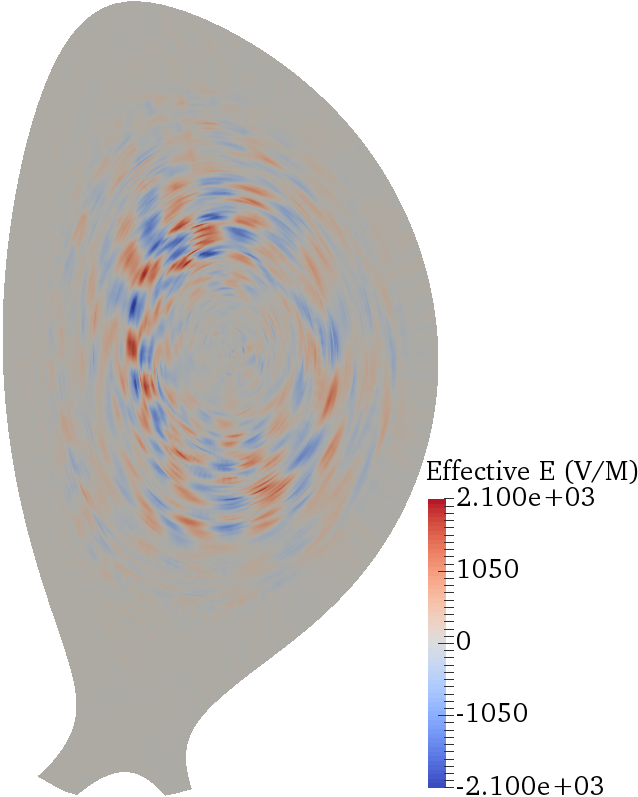} & \includegraphics[width=3.77cm, height=4.75cm]{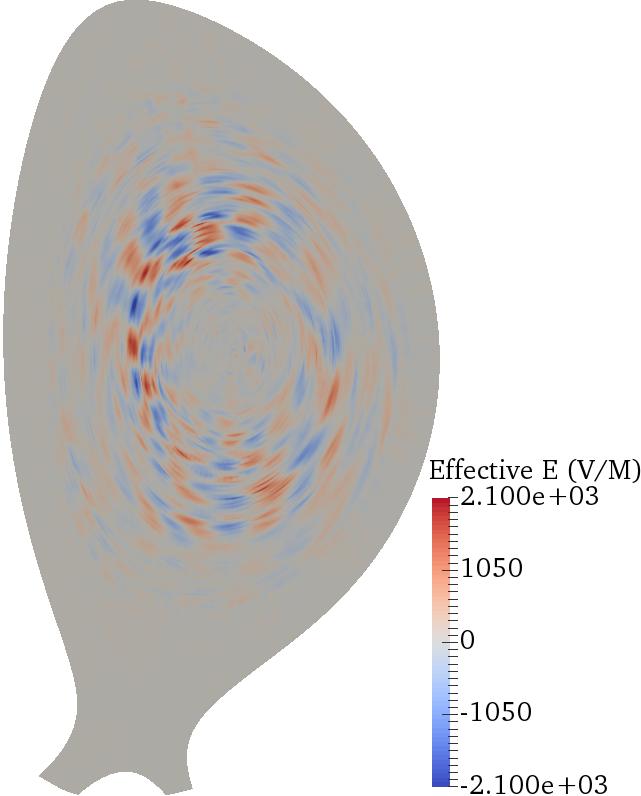} & \includegraphics[width=3.5cm, height=4.75cm]{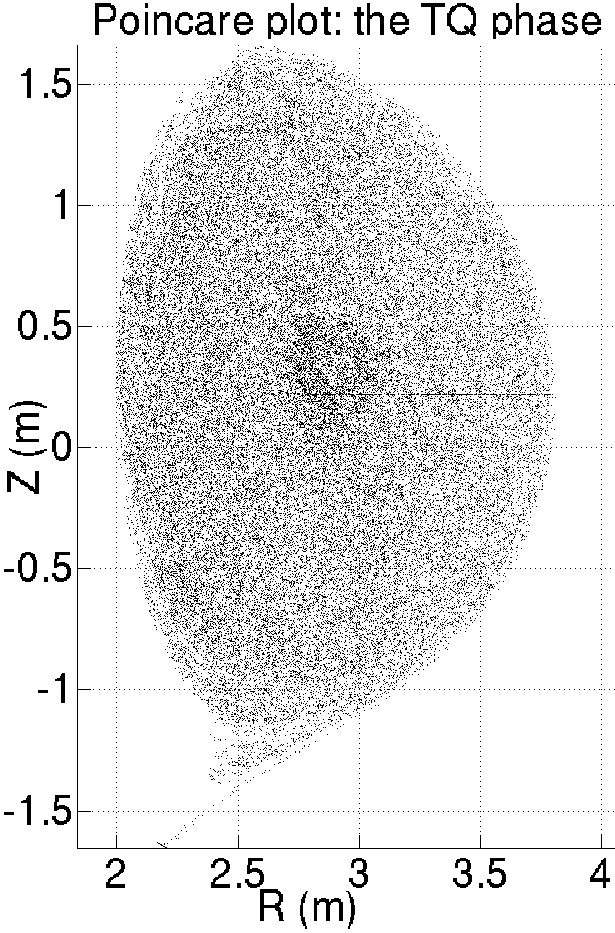} & {\begin{tabular}{c} t=4.03ms: \\ developed TQ \end{tabular}} \\
			\midrule
			\includegraphics[width=3.77cm, height=4.75cm]{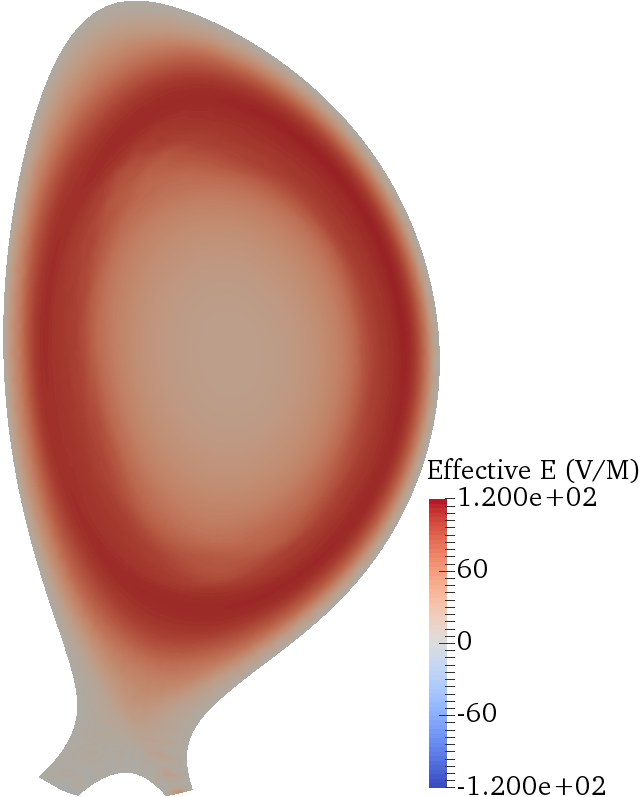} & \includegraphics[width=3.77cm, height=4.75cm]{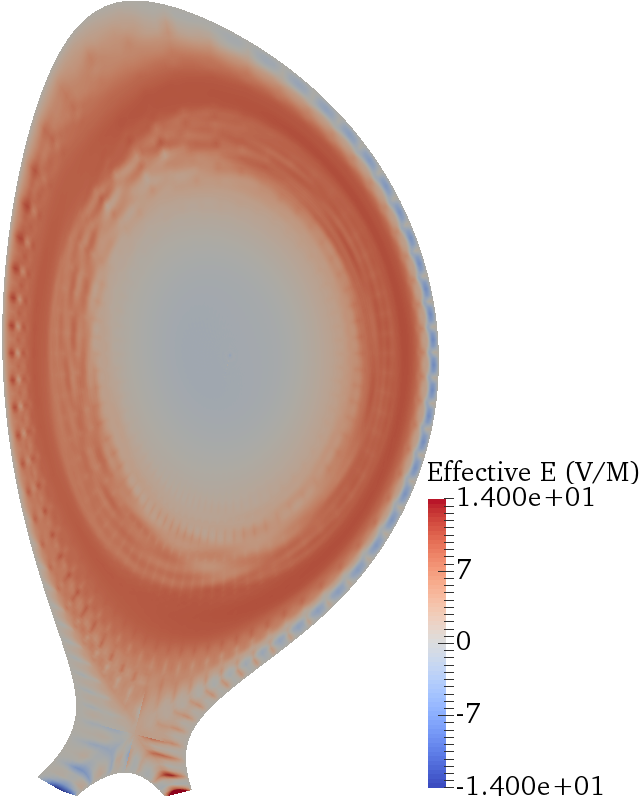} & \includegraphics[width=3.77cm, height=4.75cm]{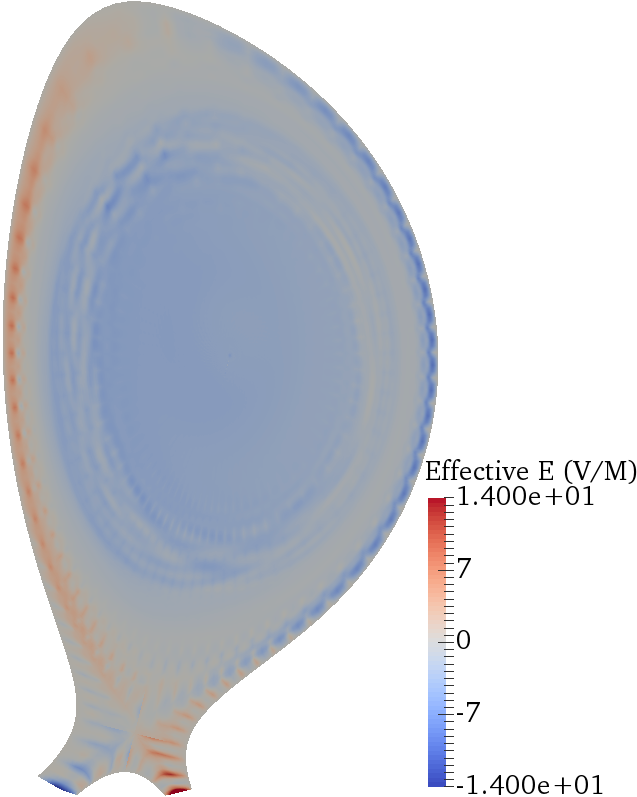}& \includegraphics[width=3.5cm, height=4.75cm]{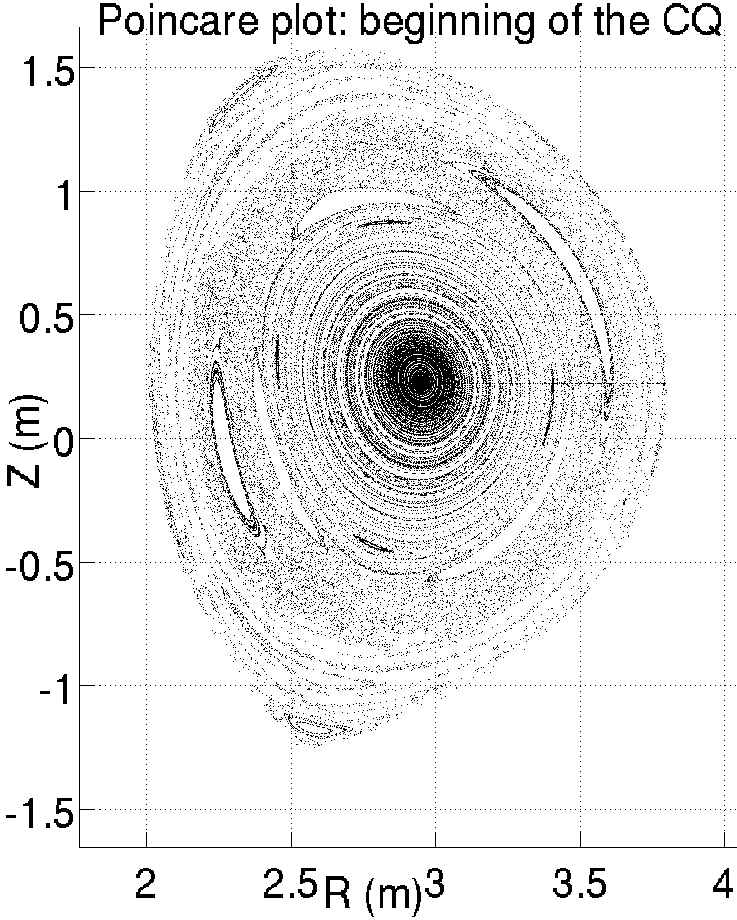} & {\begin{tabular}{c} t=6.94ms: \\ CQ beginning \end{tabular}} \\
			\bottomrule
		\end{tabular}
	\end{adjustwidth}
	\caption{Parallel effective electric field ($\mathrm{\frac{V}{m}}$) at a toroidal angle, $\phi$, of $45^{\circ}$ for a pitch angle of $170^{\circ}$. From top to bottom different disruption instants are reported: pre-TQ (t=3.55ms), TQ beginning (t=3.83ms), fully developed TQ (t=4.03ms) and CQ beginning (t=6.94ms). Kinetic energies of $\mathrm{\left[1,10,100\right]keV}$ are shown from left to right. Blue and red shades represent respectively regions of accelerating and decelerating $\mathrm{E_{\parallel,eff}}$.}
	\label{tab:chp4_Eff_phi45E1to100keVpitch170_tearingtoTQ}
\end{table}

\begin{figure}[h!]
	\centering
	\subfigure{\includegraphics[width=15cm, height=8.25cm]{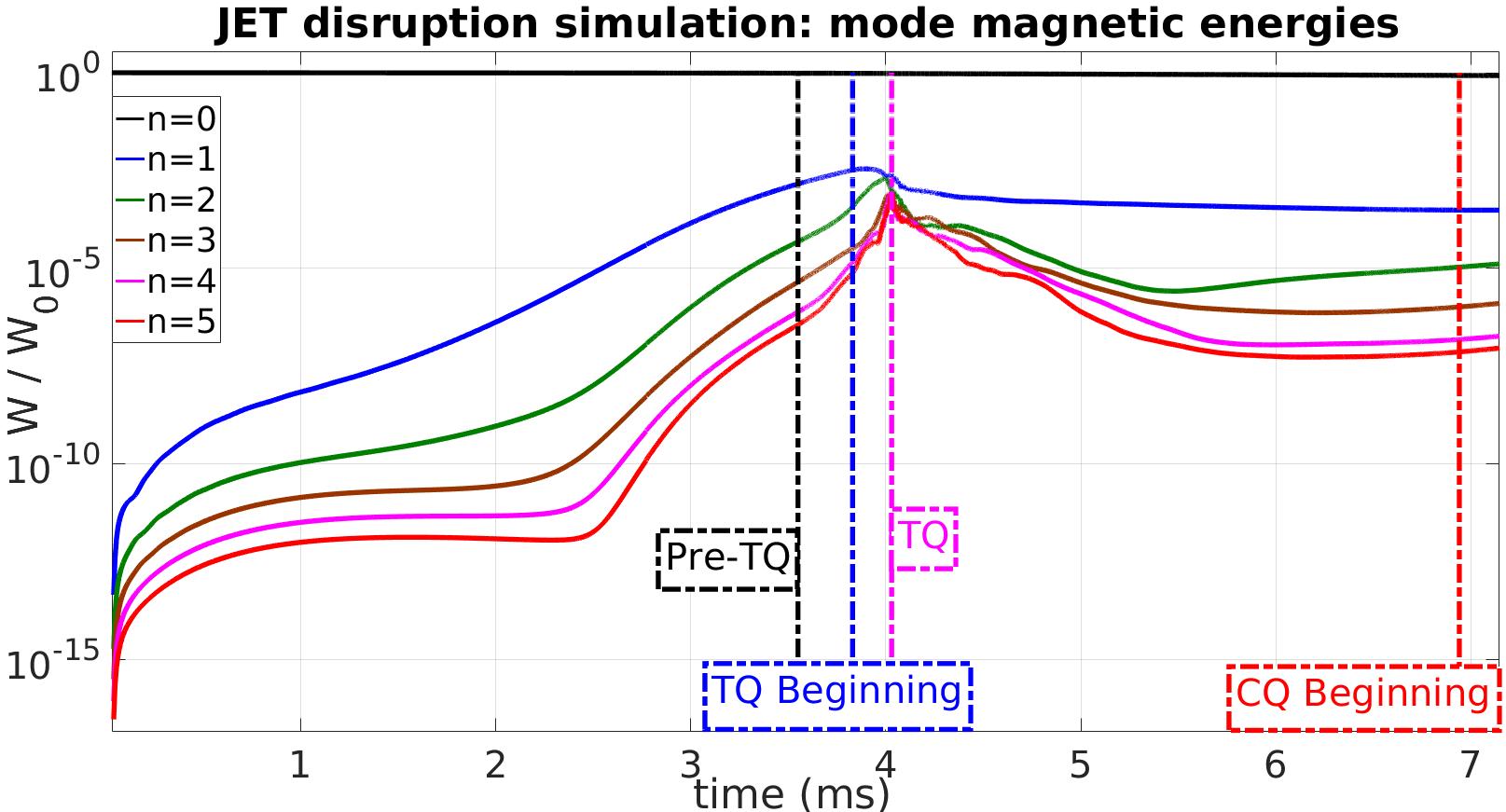}}
	\caption{Time profiles of the simulated disruption mode magnetic energies (W) normalised to the equilibrium one ($\mathrm{W_0}$). Solid lines represent the magnetic energy profiles where different colors are associated to different toroidal numbers (n). Dash-dot lines correspond to the time slices at which the $\mathrm{E_{\parallel,eff}}$ is calculated.}
	\label{fig:chp3_magnetic_energies_time_slice}
\end{figure}

\begin{figure}[h!]
	\centering
	\subfigure{\includegraphics[width=4.5cm, height=6cm]{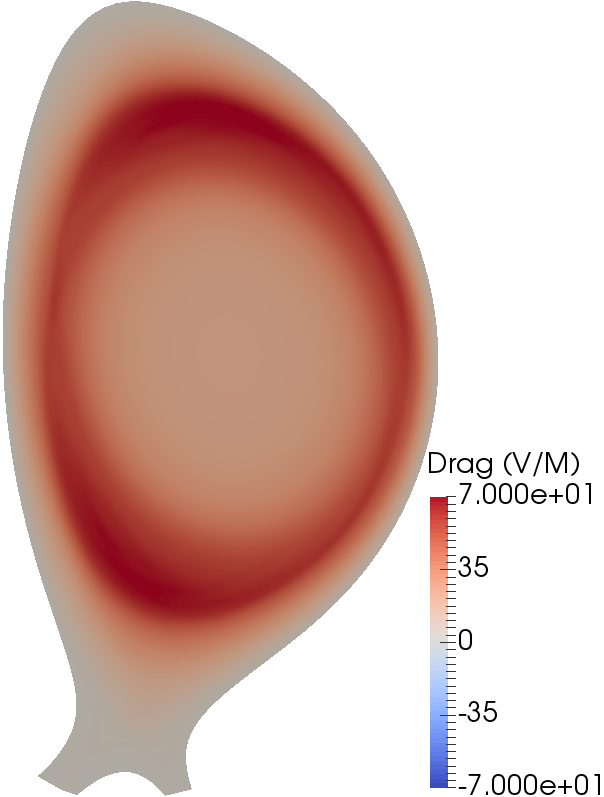} \label{fig:chp4_collDrag_phi45E1to100keVpitch170_preTQ}}
	\subfigure{\includegraphics[width=4.75cm, height=6cm]{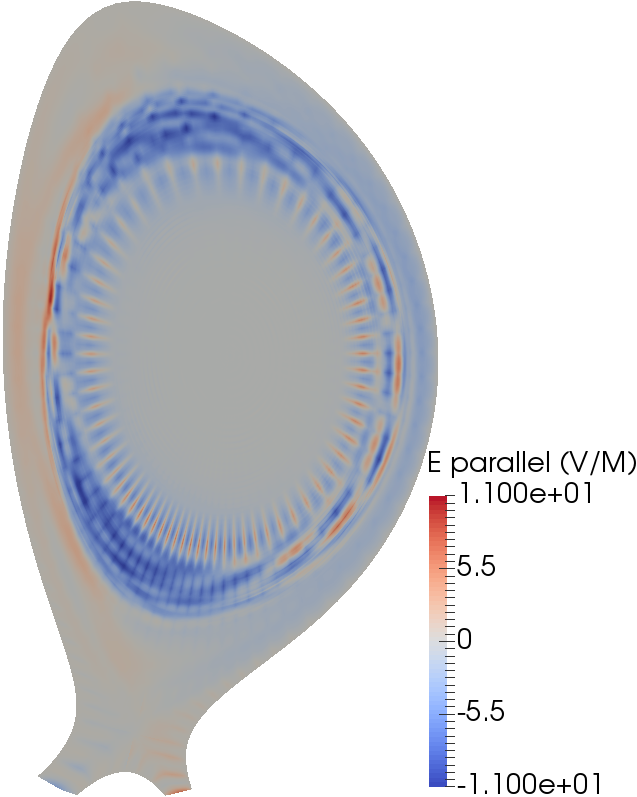} \label{fig:chp4_Efield_phi45_preTQ}}
	\caption{Decomposition of the $\mathrm{E_{kin}=1keV}$ ($\mathrm{\frac{V}{m}}$) during the pre-TQ phase (t=3.55ms): left and right plots report respectively the collision drag and the $\mathrm{E_{\parallel}}$ at $\mathrm{\phi=45^{\circ}}$. Blue and red colours represent respectively regions of accelerating and decelerating field.} 
	\label{fig:chp4_Effdecompostion_phi45_preTQ}
\end{figure}

A column-wise reading of Tables \ref{tab:chp4_Eff_phi45E1to100keVpitch170_tearingtoTQ} shows the $\mathrm{E_{\parallel,eff}}$ evolution for the times: $\mathrm{[3.55,3.83,4.03,6.94]ms}$. These time slices correspond respectively to the pre-TQ, TQ beginning, TQ and beginning of the CQ phases as visualised in Figure \ref{fig:chp3_magnetic_energies_time_slice} where solid lines represent the mode magnetic energies of the simulated disruption and dash-dot lines are associated to the $\mathrm{E_{\parallel,eff}}$ time slices. On the other hand, a row-wise scan allows comparisons between different energies. A Poincar\'e plot is also provided for each time. As discussed above, in JET the plasma current ($\mathrm{I_p}$) and the toroidal magnetic field are in the same direction while RE move always in the opposite direction \cite{papp111}. Thus, regions of negative (in blue) and positive (in red) $\mathrm{E_{\parallel,eff}}$ cause respectively the acceleration or deceleration of runaways. Let us describe the temporal evolution of $\mathrm{E_{\parallel,eff}}$. The first row of Table \ref{tab:chp4_Eff_phi45E1to100keVpitch170_tearingtoTQ} shows $\mathrm{E_{\parallel,eff}}$ when the presence of neutral gas has significantly destabilised an $\mathrm{m=2,n=1}$ tearing mode (magnetic island) but before the TQ onset. A comparison among Figures at different energies reveals that $\mathrm{E_{\parallel,eff}}$ evolves from a fully decelerating to an almost fully accelerating condition when the kinetic energy is augmented from 1keV to 100keV. This evolution is related to the reduction of collisionality with the kinetic energy increase. As remarked in \cite{nardon17}, the highest electron density rise is localised within the $\mathrm{m=2,n=1}$ magnetic island, which causes an inevitable increase of drag force and plasma resistivity in this region. At higher kinetic energies, electron collisions become less probable implying an inevitable decrease of the drag force. At the same time, the 100keV plot reveals the presence of an accelerating electric field mainly due to the increase of plasma resistivity. This double effect due to the MGI deposition in the $\mathrm{m=2,n=1}$ tearing mode is more clearly visualised in Figure \ref{fig:chp4_Effdecompostion_phi45_preTQ} which presents the 1keV collision drag (left plot) and the parallel electric field (right plot) for the same simulation time (3.55ms). When Figure \ref{fig:chp4_Effdecompostion_phi45_preTQ} is juxtaposed to the first row of Table \ref{tab:chp4_Eff_phi45E1to100keVpitch170_tearingtoTQ} it becomes evident that the augmentation of particle energy causes a transition from a collision to an electric field dominated $\mathrm{E_{\parallel,eff}}$.

The parallel effective electric field variation with the kinetic energy has important effects on the electron population dynamics. If electrons are initialised in thermal conditions ($\mathrm{E_{kin} \simeq 1keV}$) the collision drag will prevent their acceleration. On the other hand, electrons from the far tail of the thermal distribution ($\mathrm{E_{kin} \geq 10keV}$) may run away before the TQ, especially inside the $\mathrm{m=2,n=1}$ magnetic island.

We now turn our attention to the second row of Table \ref{tab:chp4_Eff_phi45E1to100keVpitch170_tearingtoTQ}, which corresponds to the beginning of the magnetic field stochastisation (in the following, we will refer to this time instant as the TQ beginning). In contrast with the previous phase, at the TQ beginning the parallel effective electric field is completely dominated by the $\mathrm{E_\parallel}$ term: the figures composing the second row of Table \ref{tab:chp4_Eff_phi45E1to100keVpitch170_tearingtoTQ} are indeed virtually impossible to distinguish. In addition, it has to be remarked that the parallel electric field activity is mainly focused at the plasma core with an intensity two orders of magnitude higher than at t=3.55ms (i.e. during the pre-TQ phase).

The third row of figures composing Table \ref{tab:chp4_Eff_phi45E1to100keVpitch170_tearingtoTQ} is dedicated to the $\mathrm{E_{\parallel,eff}}$ acting during the MHD-activity peak of the TQ, when closed magnetic surfaces are completely destroyed. The characterisation of the source and type of the observed MHD fluctuations is beyond the scope of this paper but, at this stage, a plausible hypothesis involves the generation of MHD turbulence by the magnetic field ergodisation as discussed in \cite{biskamp97} Chapter 8.2.5. Further theoretical works on the subject are reported in \cite{biskamp84} and in \cite{diamond84} but they do not describe the $\mathrm{E_{\parallel}}$ evolution during the TQ. As before, solutions for $\mathrm{\left[1,10,100\right]keV}$ do not differ significantly so, also in this case, the $\mathrm{E_{\parallel}}$ term is dominant. The strongest electric activity is around mid-radius which is where $\mathrm{E_{\parallel,eff}}$ fluctuations up to $\sim$2kV are observed. During this phase, the electric field presents a cellular-like topology with an alternation of accelerating and decelerating regions in the poloidal direction which extends up to the plasma edge. These cells are smaller than the ones observed at the TQ beginning (t=3.83ms) but their intensities are similar. This particular $\mathrm{E_{\parallel,eff}}$ topology suggests that it is mainly due to the large MHD fluctuations taking place during this phase but a dedicated analysis would be required to understand the precise mechanisms at play.

\begin{figure}[h!]
	\centering
	\subfigure{\includegraphics[width=5.5cm, height=6cm]{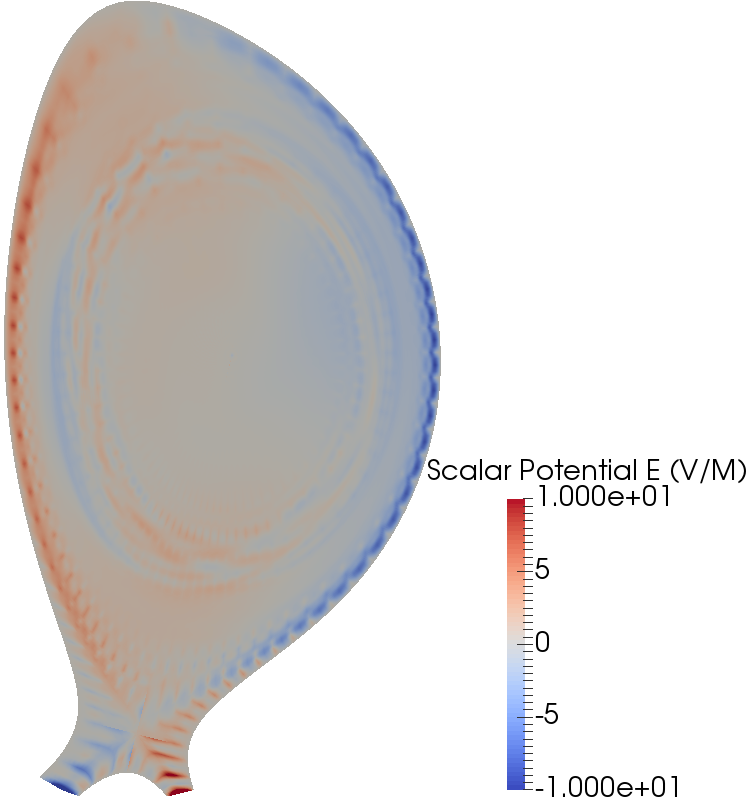} \label{fig:chp4_Epotential_phi45_BeginningCQ}}
	\subfigure{\includegraphics[width=5.5cm, height=6cm]{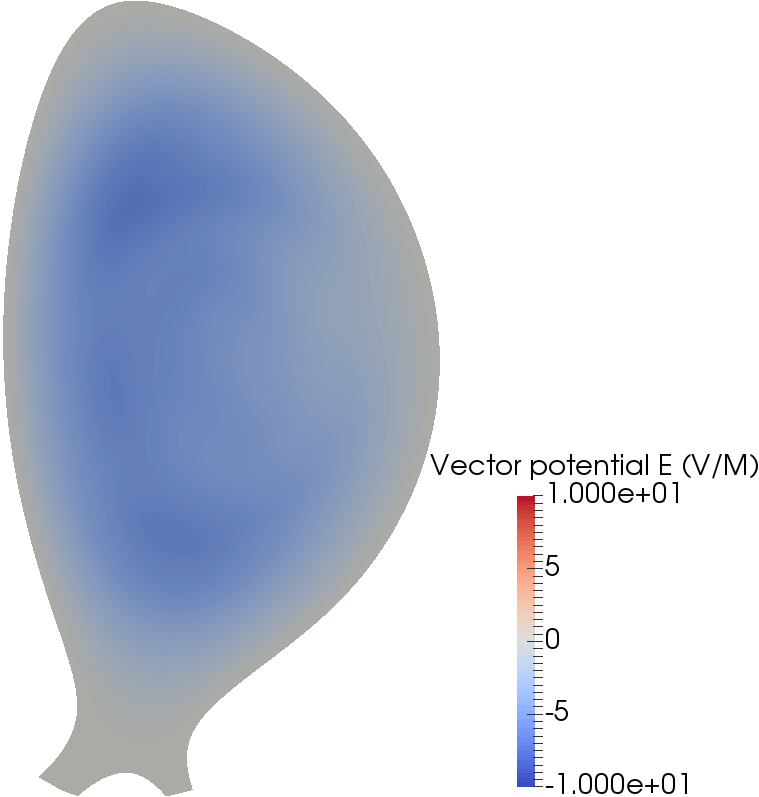} \label{fig:chp4_Edpsidt_phi45_BeginningCQ}}
	\caption{Electric scalar (left) and vector (right) potential contributions to the $\mathrm{E_{\parallel}}$ ($\mathrm{\frac{V}{m}}$) at the beginning of the CQ (t=6.94ms). Blue and red colours represent respectively regions of accelerating and decelerating field} 
	\label{fig:chp4_Epardecomposition_phi45_BeginningCQ}
\end{figure}

The last row of Table \ref{tab:chp4_Eff_phi45E1to100keVpitch170_tearingtoTQ} corresponds to the beginning of the CQ (t=6.94ms) which is characterised by the presence of large areas having good confinement properties and by the beginning of the plasma current decay. The CQ stage is distinguished by the return of the competition between collision drag and accelerating electric field: $\mathrm{1keV}$ electrons are always decelerated due to high collision braking while particles having a kinetic energy $\mathrm{\geq 10keV}$ and confined in the plasma core will be accelerated and become runaway. As remarked for the 1keV plot of the Table \ref{tab:chp4_Eff_phi45E1to100keVpitch170_tearingtoTQ} first row, the drag force is stronger at the plasma edge due to the higher MGI-induced increase in electron density. In Figure \ref{fig:chp4_Epardecomposition_phi45_BeginningCQ} the parallel electric field is decomposed into scalar and vector potential components. Figure \ref{fig:chp4_Epardecomposition_phi45_BeginningCQ} shows that the $\mathrm{E_{\parallel}}$ is dominated by the $\mathrm{\frac{\partial \psi}{\partial t}}$ (inductive) term, which is related to the plasma current decay caused by the increase of plasma resistivity.
%
%The second row of Table \ref{tab:chp4_Eff_phi45E1to100keVpitch170_endTQtoBeginCQ} corresponds to the beginning of the CQ (t=6.94ms) which is characterised by the presence of large areas having good confinement properties and by the beginning of the plasma current decay. The CQ stage is distinguished by the return of the competition between collision drag and accelerating electric field: $\mathrm{1keV}$ electrons are always decelerated due to high collision braking while particles having a kinetic energy $\mathrm{\geq 10keV}$ and confined in the plasma core will be accelerated and become runaway. As remarked for the 1keV plot of the Table \ref{tab:chp4_Eff_phi45E1to100keVpitch170_tearingtoTQ} first row, the drag force is stronger at the plasma edge due to the higher MGI-induced increase in electron density. In Figure \ref{fig:chp4_Epardecomposition_phi45_BeginningCQ} the parallel electric field is decomposed into scalar and vector potential components. Figure \ref{fig:chp4_Epardecomposition_phi45_BeginningCQ} shows that the $\mathrm{E_{\parallel}}$ is dominated by the $\mathrm{\frac{\partial \psi}{\partial t}}$ (inductive) term, which is related to the plasma current decay caused by the increase of plasma resistivity.

At this point of our discussion, we present a summary of Table \ref{tab:chp4_Eff_phi45E1to100keVpitch170_tearingtoTQ} results. In the pre-TQ and CQ phases (respectively first and last rows of Table \ref{tab:chp4_Eff_phi45E1to100keVpitch170_tearingtoTQ}), $\mathrm{E_{\parallel,eff}}$ is strongly dependent on the kinetic energy. The dynamics of a 1keV (thermal) electron is everywhere dominated by the drag force; thus, a thermal population cannot reach runaway energies in these time periods. Conversely for kinetic energies of 10keV, regions of accelerating $\mathrm{E_{\parallel}}$ appear at the plasma core allowing the generation of RE. Further increase of $\mathrm{E_{kin}}$ implies a greater drag force reduction thus stronger accelerating electric fields which extend towards the plasma edge. In contrast, all along the TQ (second and third rows of Table \ref{tab:chp4_Eff_phi45E1to100keVpitch170_tearingtoTQ}), the most prominent contribution to $\mathrm{E_{\parallel,eff}}$ is given by $\mathrm{E_{\parallel}}$. During this phase, the MHD activity generates cells of accelerating and decelerating electric fields which strengthen and reduce in size until the complete magnetic field stochastisation (t=4.03ms) and then decay. The presence of large $\mathrm{E_{\parallel,eff}}$ fluctuations at the TQ and the possibility to generate RE if $\mathrm{E_{kin} >10keV}$ during the CQ raise the question of whether the TQ electric fields are able to accelerate a fraction of an initially thermal electron population up to this critical energy level leading to the formation of RE. This question motivates the study presented in Section \ref{chp4_electron_acceleration_sec_electron_accelerarion_TQ}.
\section{Electron acceleration during the TQ phase} \label{chp4_electron_acceleration_sec_electron_accelerarion_TQ}

In this Section we study the capability of the TQ electric field to accelerate a thermal electron up to runaway conditions. For this purpose, we track multiple test GC populations from t=4.01ms, i.e., just before the time of the Table \ref{tab:chp4_Eff_phi45E1to100keVpitch170_tearingtoTQ} third row, up to the beginning of the CQ (t=6.94ms), for a total simulated time of $\mathrm{\sim 3 ms}$. The late initialisation in the disruption simulation (t=4.01ms) is necessary to avoid a significant decrease of $\mathrm{p_{\parallel}}$ due to the intense drag force typical of the $\mathrm{E_{kin}=1keV}$ case. This momentum reduction is caused by the drag force model dissipation which does not preserve the thermodynamic equilibrium. The procedure is similar to the one exposed in \cite{sommariva18}: the plasma minor radius, expressed in normalised magnetic flux coordinates, is divided into 10 nodes from the plasma core to the edge. A population consisting of $10^3$ GC is randomly initialised on each n=0 surface identified by a specific normalised poloidal flux label ($\mathrm{\bar{\psi}}$) denoted $\mathrm{\bar{\psi}_{init}}$ (n is the toroidal mode number). It has to be remarked that the $\mathrm{\bar{\psi}}$ coordinate system is chosen to be the one used for the particle initialisation procedure and is kept constant all along the simulation. As done in \cite{sommariva18}, a mono-energetic mono-pitch angle electron beam is used. The chosen energies are: $\mathrm{E_{kin}=[1,5,10,25,50]keV}$ while the pitch angle is set to $\mathrm{170^{\circ}}$ (counter-current passing particle) for each run. Then, the electron distribution is evolved in the disruption simulation using a time step of $\mathrm{14{\cdot}T_{gyro}}$ (where $\mathrm{T_{gyro}}$ is the non-relativistic gyration period) which was shown to be a good compromise between result accuracy and simulation computational cost \cite{sommariva18}.   

\begin{figure}[h!]
	\centering
	%\subfigure{\includegraphics[width=10cm, height=6cm]{figure_chp_REacc/Ekin_vs_time_Ekin1keV_ngc1000_nout800_psi0e05_pitch170_eta1en7_zoom.png} \label{fig:chp4_EkinProf_psi0e05_Ekin1keV_pitch170_tot}}
	%\subfigure{\includegraphics[width=10cm, height=6cm]{figure_chp_REacc/ppar_vs_time_Ekin1keV_ngc1000_nout800_psi0e05_pitch170_eta1en7_zoom.png} \label{fig:chp4_pparProf_psi0e05_Ekin1keV_pitch170_TQ}}
	\subfigure{\includegraphics[width=10cm, height=5.8cm]{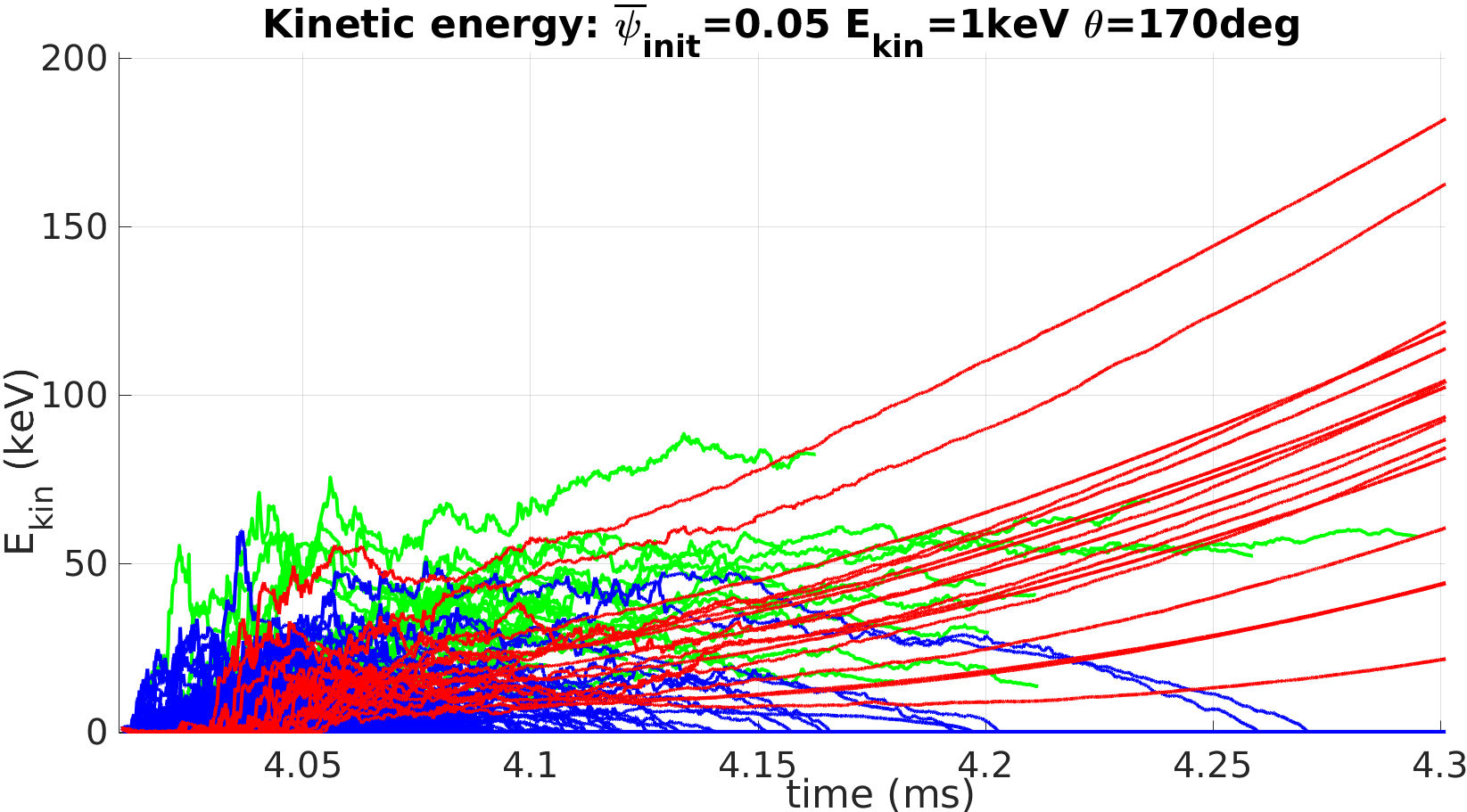} \label{fig:chp4_EkinProf_psi0e05_Ekin1keV_pitch170_tot}}
	\subfigure{\includegraphics[width=10cm, height=5.8cm]{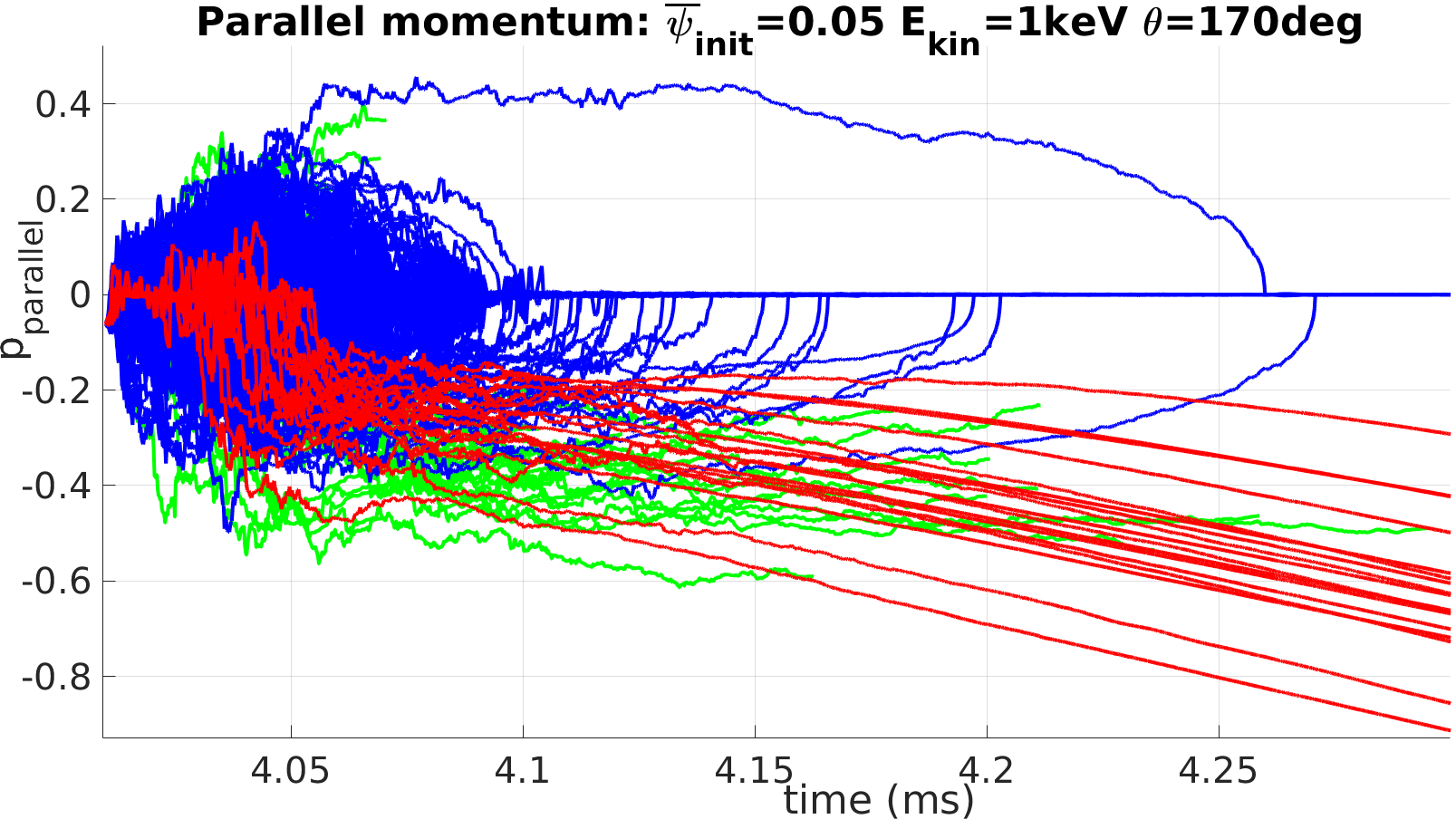} \label{fig:chp4_pparProf_psi0e05_Ekin1keV_pitch170_TQ}}
	\caption{Kinetic energy (upper plot) and parallel momentum (lower plot) time profiles for a population initialised in the plasma core region with a kinetic energy of $\mathrm{1keV}$ at the early phase of the disruption simulation. Green lines are lost particles while red and blue lines are respectively particles having final energy above and below 1MeV.}
	\label{fig:chp4_EkinPparProf_psi0e05_Ekin1keV_pitch170}
\end{figure}

In Figure \ref{fig:chp4_EkinPparProf_psi0e05_Ekin1keV_pitch170}, the kinetic energy (upper plot) and the parallel momentum (lower plot) time profiles for $\mathrm{10^3}$ particles initialised in the core region with an initial kinetic energy of $\mathrm{1keV}$ are displayed for the first $\mathrm{0.3ms}$ of the simulation. Lost particle profiles are shown using green lines while red and blue lines are associated to electrons having a final energy respectively above and below $\mathrm{1MeV}$. It has to be noted that this set of initial conditions is representative of a core background electron population. Indeed, the particle initial kinetic energy is consistent with the background thermal one at this time of the simulated disruption. Moreover, the core magnetic field is ergodised just before the TQ MHD activity peak therefore, core electrons remains confined during both the pre-TQ and the TQ beginning phases justifying the late particle initialisation discussed in Section \ref{chp4_sec_collisional_drag}. The first plot of Figure \ref{fig:chp4_EkinPparProf_psi0e05_Ekin1keV_pitch170} clearly shows that a fraction of the initial population (in blue) loses its kinetic energy until reaching the minimum energy level allowed by the drag operator. Contrarily, a few electrons (in red) see an increase in their kinetic energies up to relativistic conditions, positively answering to the question asked at the end of Section \ref{chp4_sec_electric_field_disruption}. Recalling the interpretation also given in Section \ref{chp4_sec_electric_field_disruption}, $\mathrm{1keV}$ particles which are not accelerated during the TQ cannot reach runaway conditions during the CQ because, for this energy level, the collision drag overwhelms the driving electric field at each radial position (last row-left plot of Table \ref{tab:chp4_Eff_phi45E1to100keVpitch170_tearingtoTQ}). This means that the simulated runaway electrons interact with regions of counter-$\mathrm{I_p}$ accelerating $\mathrm{E_{\parallel,eff}}$ during the TQ. Despite the beam-type initialisation, which obliges all electrons to have equal parallel momentum, when the population enters the TQ phase a significant spread of the distribution function in velocity space is recorded. This spreading is mainly caused by the presence of accelerating and decelerating electric cells shown in the second and third rows of Table \ref{tab:chp4_Eff_phi45E1to100keVpitch170_tearingtoTQ}.

The intense $\mathrm{p_{\parallel}}$ fluctuations in concert with the permanent presence of particles at the plasma core (discussed in \cite{sommariva18}) allow the confinement of electrons (less than $\mathrm{2\%}$ of the initial population) having energies high enough and the correct direction ($\mathrm{p_{\parallel}<0}$) for becoming runaway due to the CQ inductive electric field, as depicted by the red lines of Figure \ref{fig:chp4_EkinPparProf_psi0e05_Ekin1keV_pitch170}.

\begin{figure}[h!]
	\centering
	\includegraphics[width=10cm, height=6.3cm]{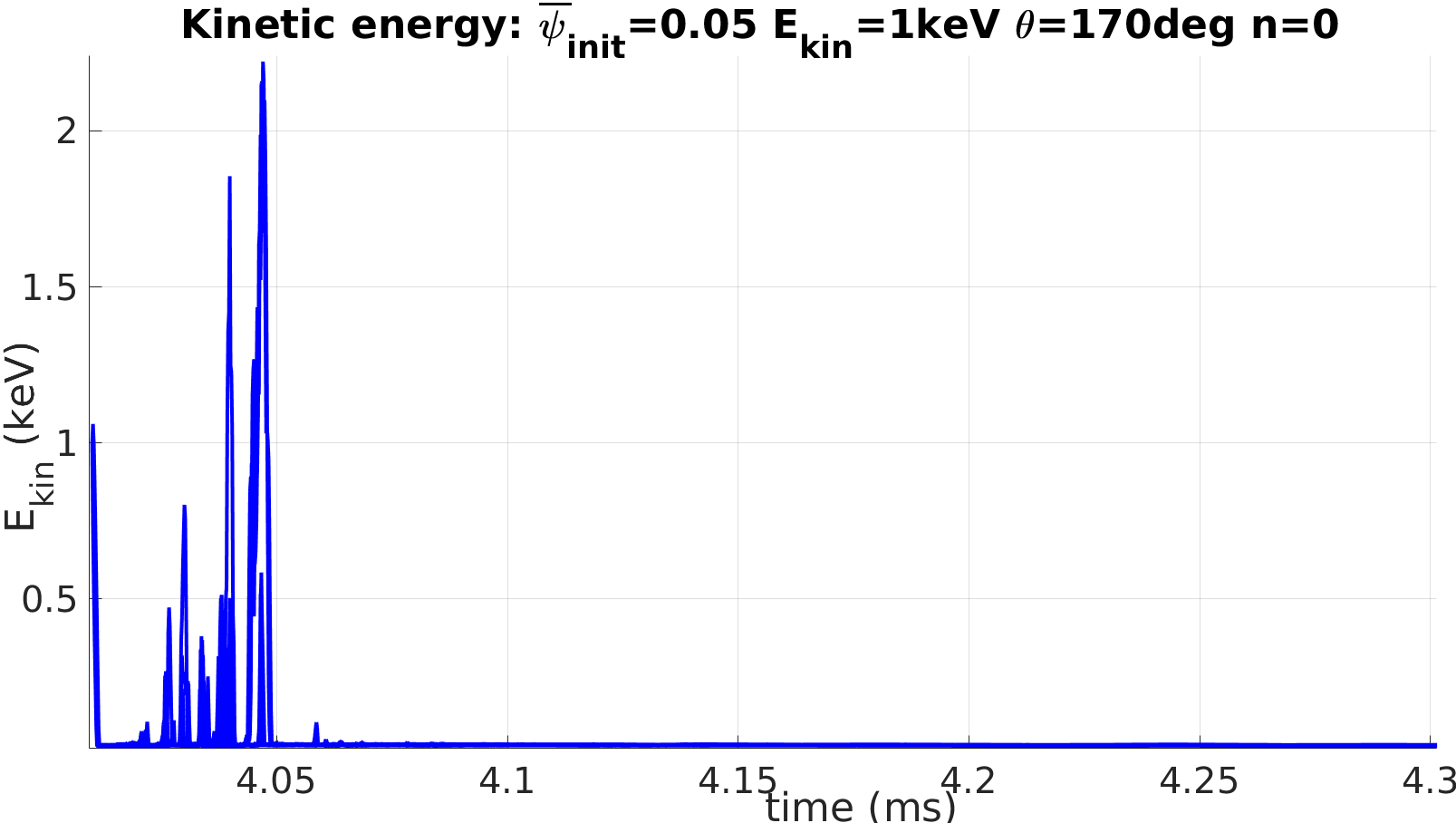} \label{fig:chp4_EkinProf_psi0e05_Ekin1keV_pitch170_tot}
	\caption{Kinetic energy time profiles for an electron population (initialised at the plasma core with a kinetic energy of $\mathrm{1keV}$) when the plasma fields are described using only the n=0 toroidal harmonic. It is worth remarking that neither lost nor runaway electrons are observed in this simulation.}
	\label{fig:chp4_EkinPparProf_psi0e05_Ekin1keV_pitch170_neq0}
\end{figure}

In order to assess the importance of MHD fluctuations in the electron acceleration process described above, the same simulation presented in Figure \ref{fig:chp4_EkinPparProf_psi0e05_Ekin1keV_pitch170} was conducted using only the n=0 component of the background plasma fields. The electron kinetic energy time traces for this simulation are reported in Figure \ref{fig:chp4_EkinPparProf_psi0e05_Ekin1keV_pitch170_neq0}. 

From Figure \ref{fig:chp4_EkinPparProf_psi0e05_Ekin1keV_pitch170_neq0} it is clear that the electron acceleration during the TQ is strictly related to the presence of MHD fluctuations indeed, when only the n=0 mode is used the maximum kinetic energy variation is a hundred times smaller than the one obtained using the full background plasma fields, preventing the generation of REs.

\begin{figure}[h!]
	\centering
	\subfigure{\includegraphics[width=5cm, height=5.6cm]{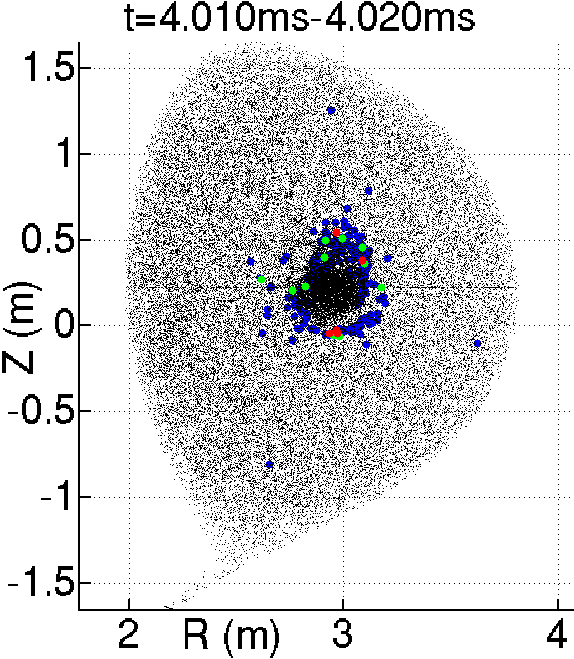} \label{chp4_pseuPoincare_plot_psi0e05_Ekin1keV_rst4000}}
	\subfigure{\includegraphics[width=5cm, height=5.6cm]{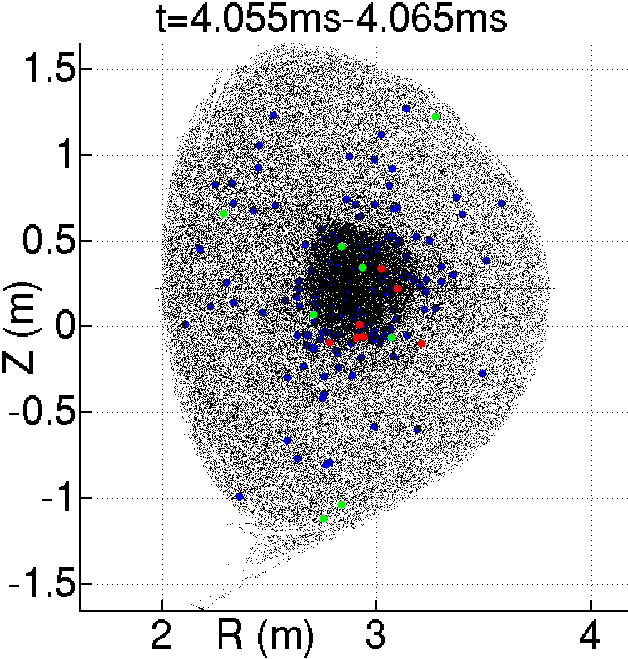} \label{chp4_pseuPoincare_plot_psi0e05_Ekin1keV_rst4490}}
	\subfigure{\includegraphics[width=5cm, height=5.6cm]{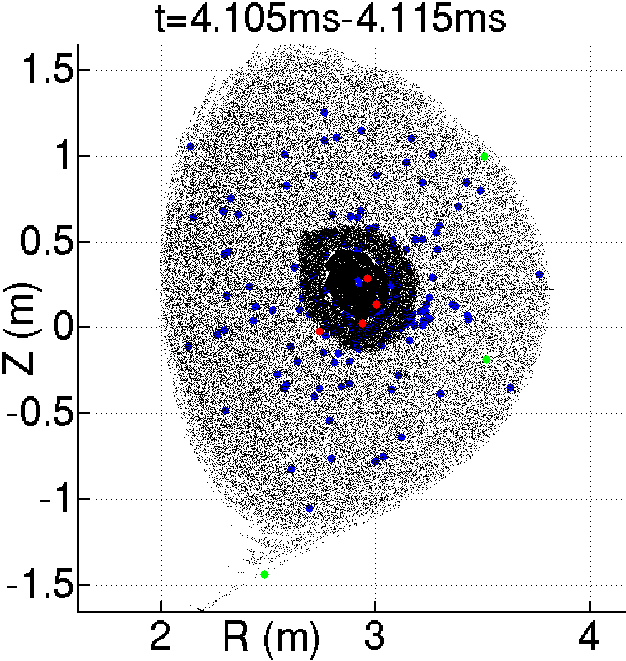} \label{fig:chp4_pseuPoincare_plot_psi0e05_Ekin1keV_rst4920}}
	\subfigure{\includegraphics[width=5cm, height=5.6cm]{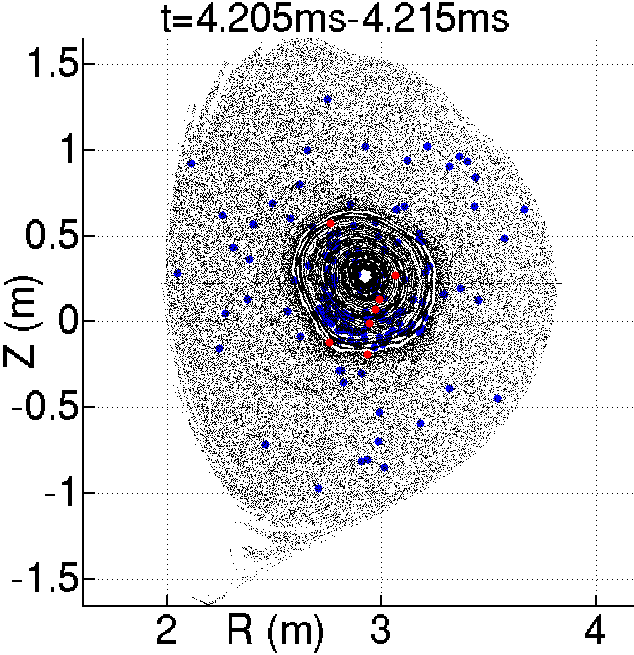} \label{fig:chp4_pseuPoincare_plot_psi0e05_Ekin1keV_rst5490}}
	\caption{Pseudo-Poincar\'e plots ($\mathrm{\phi=180^{\circ}}$) for a 1keV population initialised at the plasma core ($\mathrm{\bar{\psi}_{init}=0.05}$), for different simulation times: t=0.005ms (top left), t=0.05ms (top right), t=0.1ms (bottom left) and t=0.2ms (bottom right). Red, blue and green dots correspond respectively to electrons having final $\mathrm{E_{kin}\geq 1MeV}$, final $\mathrm{E_{kin} < 1MeV}$ and lost particles. Black dots represent the field line positions.}
	\label{fig:chp4_pseudopoincares_phi180_dphi30_psi0e05_Ekin1keV_rst4000_5490}
\end{figure}

Figure \ref{fig:chp4_pseudopoincares_phi180_dphi30_psi0e05_Ekin1keV_rst4000_5490} reports the electron Pseudo-Poincar\'e and the field line Poincar\'e plots for the times: t=4.015ms (top left), t=4.06ms (top middle), t=4.11ms (top right) and t=4.21ms (bottom middle). Red, blue and green dots denote respectively electrons having final kinetic energy above 1MeV, below 1MeV and lost particles while field line positions are identified with black dots. As introduced in \cite{sommariva18}, Pseudo-Poincar\'e plots represent the nearest particle positions to a specific poloidal plane within a given time interval. Figure \ref{fig:chp4_pseudopoincares_phi180_dphi30_psi0e05_Ekin1keV_rst4000_5490} is obtained using the $\mathrm{\phi=180^{\circ}}$ plane as reference, a toroidal angle interval of $\mathrm{\pm30^{\circ}}$ and a time window of $\mathrm{\pm0.005ms}$. As observed in \cite{sommariva18}, the magnetic stochasticity destroys the initial particle torus (Figure \ref{fig:chp4_pseudopoincares_phi180_dphi30_psi0e05_Ekin1keV_rst4000_5490} top left plot) spreading electrons in the whole plasma volume (Figure \ref{fig:chp4_pseudopoincares_phi180_dphi30_psi0e05_Ekin1keV_rst4000_5490} top middle plot). In this simulation, electrons dispersed outside the plasma centre do not reach runaway energies. Indeed, they are either lost to the wall (green dots) or decelerated to low kinetic energies (blue dots) probably due to the MGI-induced increase of collisionality in the plasma outer region. Conversely, those remaining close to the magnetic axis are reconfined by the reformation of magnetic surfaces at the end of the TQ chaotic phase. In this case, electrons being accelerated before reconfinement become RE (red dots) otherwise they are slowed down by collisions (blue dots). Thereby, the probability of an electron to become runaway seems to be related to the combined effects of the momentum and physical space transports.

\begin{figure}[h!]
	\centering
	\subfigure{\includegraphics[width=12cm, height=6.25cm]{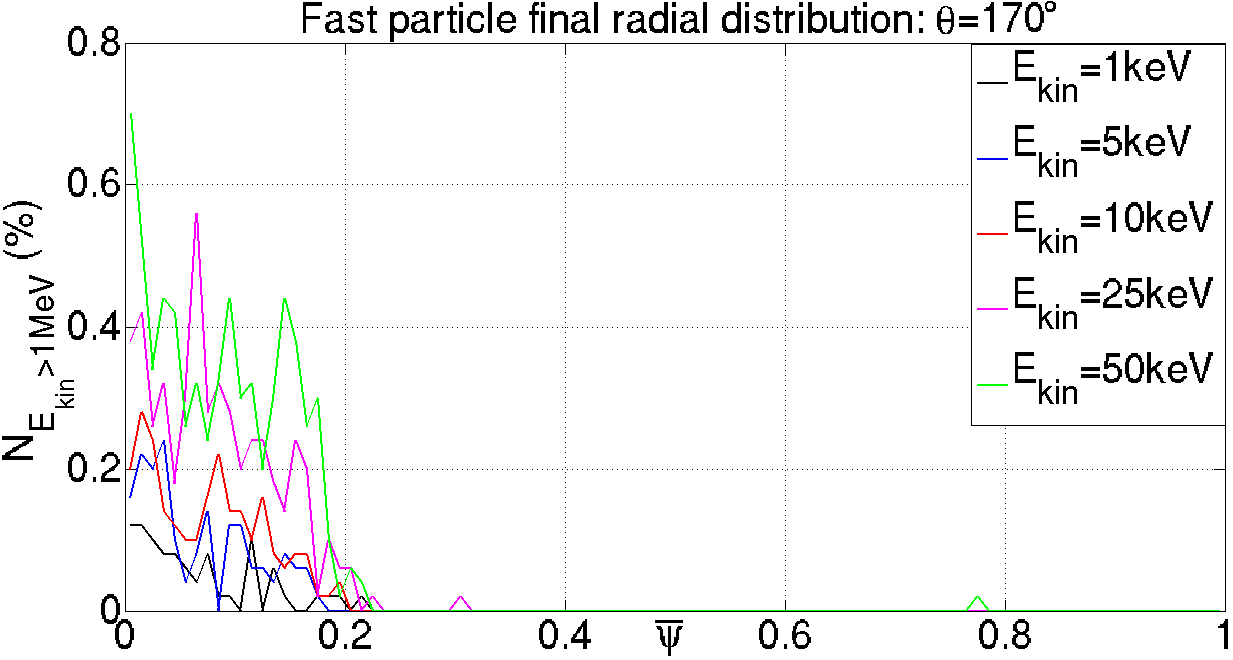}}
	\caption{Radial histogram at t=6.94ms of electrons having $\mathrm{E_{kin}>1MeV}$ (averaged over all $\mathrm{\bar{\psi}_{init}}$). The ordinate axis represents the fraction of the initial electron population having final kinetic energy above 1MeV. Black, blue, red, magenta and green lines are respectively associated to initial kinetic energies of $\mathrm{[1,5,10,25,50]keV}$.}
	\label{fig:chp4_REfinalRadialPosition_avarage}
\end{figure}

\begin{figure}[h!]
	\centering
	\subfigure{\includegraphics[width=12.5cm, height=6.25cm]{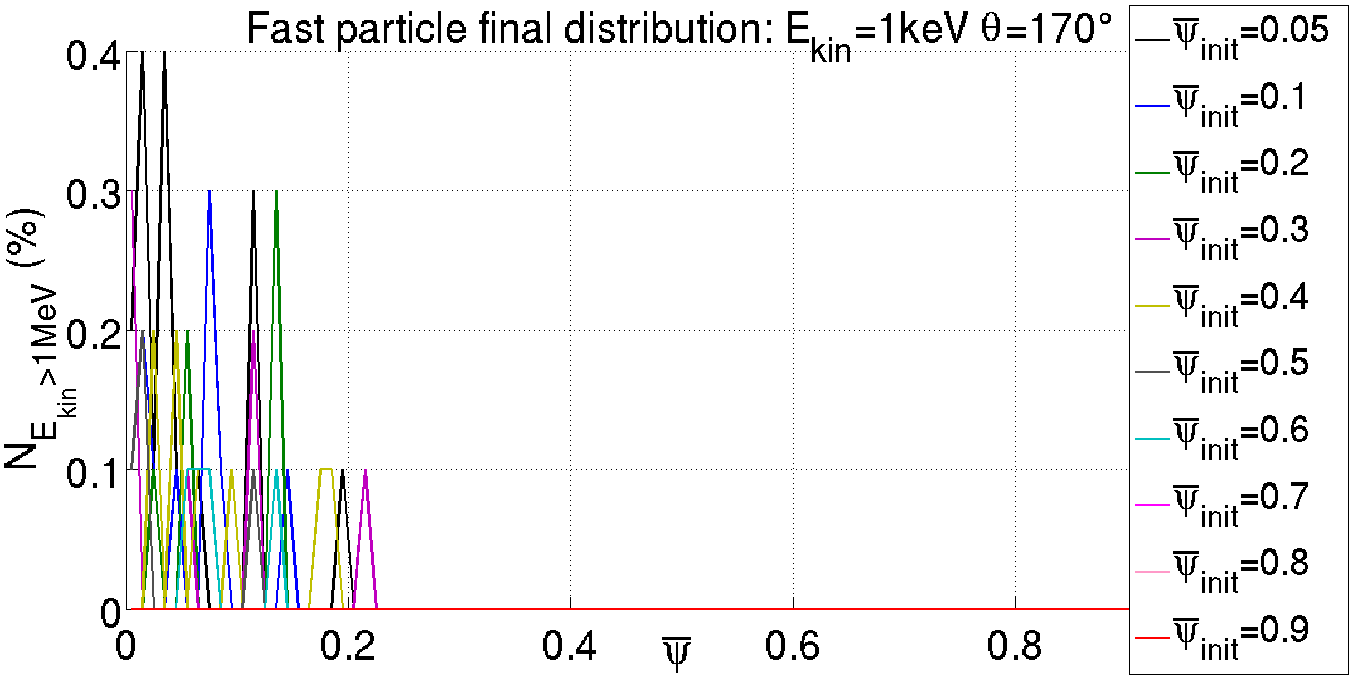}}
	\caption{Radial histogram at t=6.94ms of electrons having $\mathrm{E_{kin}>1MeV}$ and being initialised at $\mathrm{E_{kin}=1keV}$. The ordinate axis represents the fraction of the initial electron population having final kinetic energy above 1MeV. Each line colour is associated to a specific initial radial position.}
	\label{fig:chp4_REfinalRadialPosition_1keV}
\end{figure}

Figures \ref{fig:chp4_REfinalRadialPosition_avarage} and \ref{fig:chp4_REfinalRadialPosition_1keV} report the RE final radial distribution for a set of initial kinetic energies and initial radial positions respectively.

Figure \ref{fig:chp4_REfinalRadialPosition_avarage} furnishes the final RE radial distribution averaged over all initial radii in terms of normalised poloidal flux, for different initial kinetic energies. Clearly, RE are focused in the plasma core region ($\bar{\psi} \leq 0.2$). This beam-like focusing happens independently from the particle initial position as shown in Figure \ref{fig:chp4_REfinalRadialPosition_1keV} for the 1keV case.

The RE near-to-magnetic-axis positioning found in our simulation seems to be in agreement with observations obtained from DIII-D \cite{hollmann13} and, less clearly, from TEXTOR \cite{abdullaev16} RE experiments, where the measured RE synchrotron radiation suggest that the beam is mainly localised at the core of the post-disruptive plasma.

As last part of this Section, the dependencies between the total number of generated RE and the electron population initial radial position and $\mathrm{E_{kin}}$ are discussed.
% These data are reported in Figure \ref{fig:chp4_REgeneration_and_loss_psiinitvsEkininit}. 

\begin{figure}[h!]
	\centering
	\subfigure{\includegraphics[width=11.5cm, height=6.25cm]{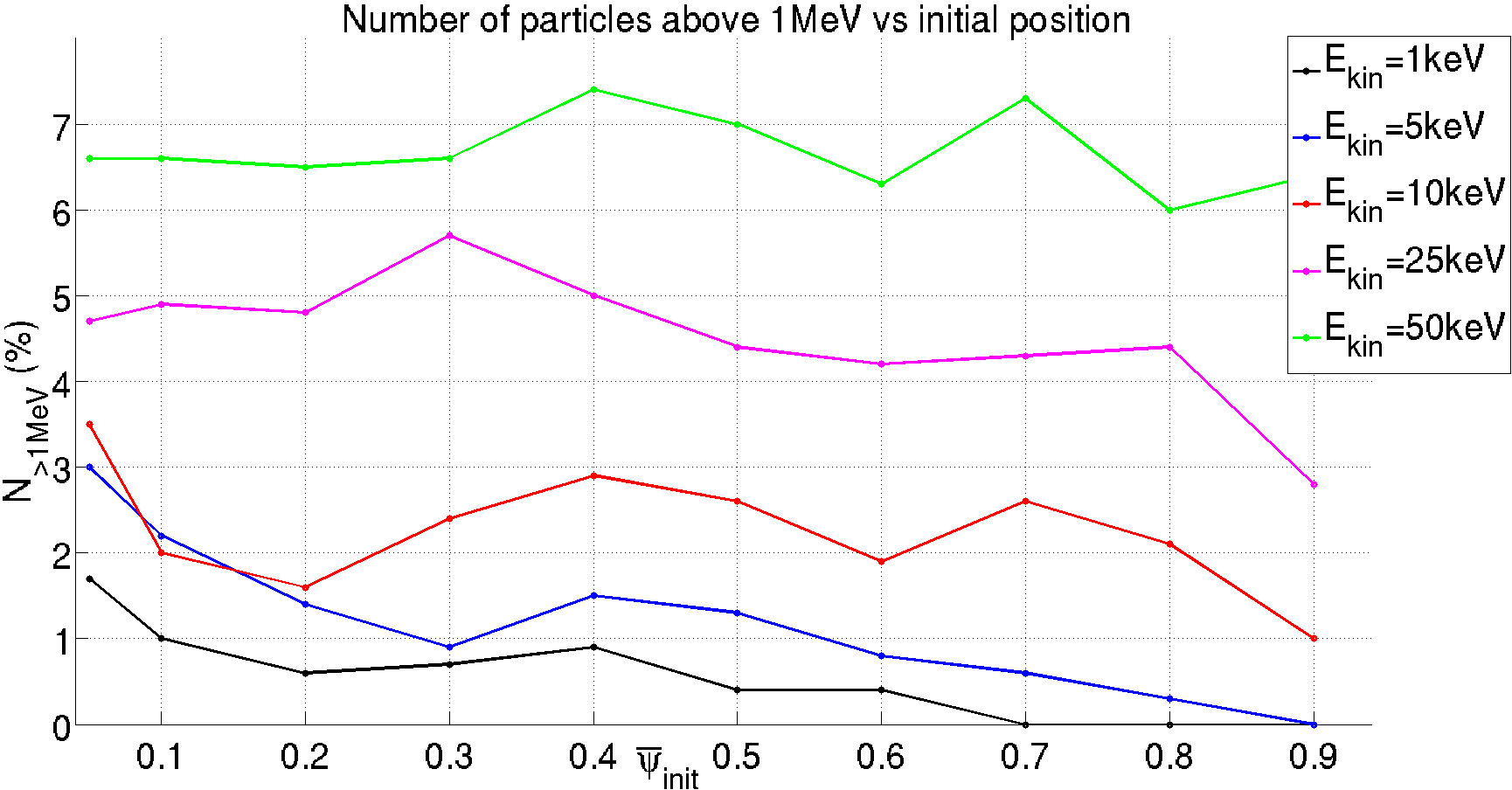} \label{fig:chp4_REgeneration_psiinitvsEkininit}}
	\subfigure{\includegraphics[width=11.5cm, height=6.25cm]{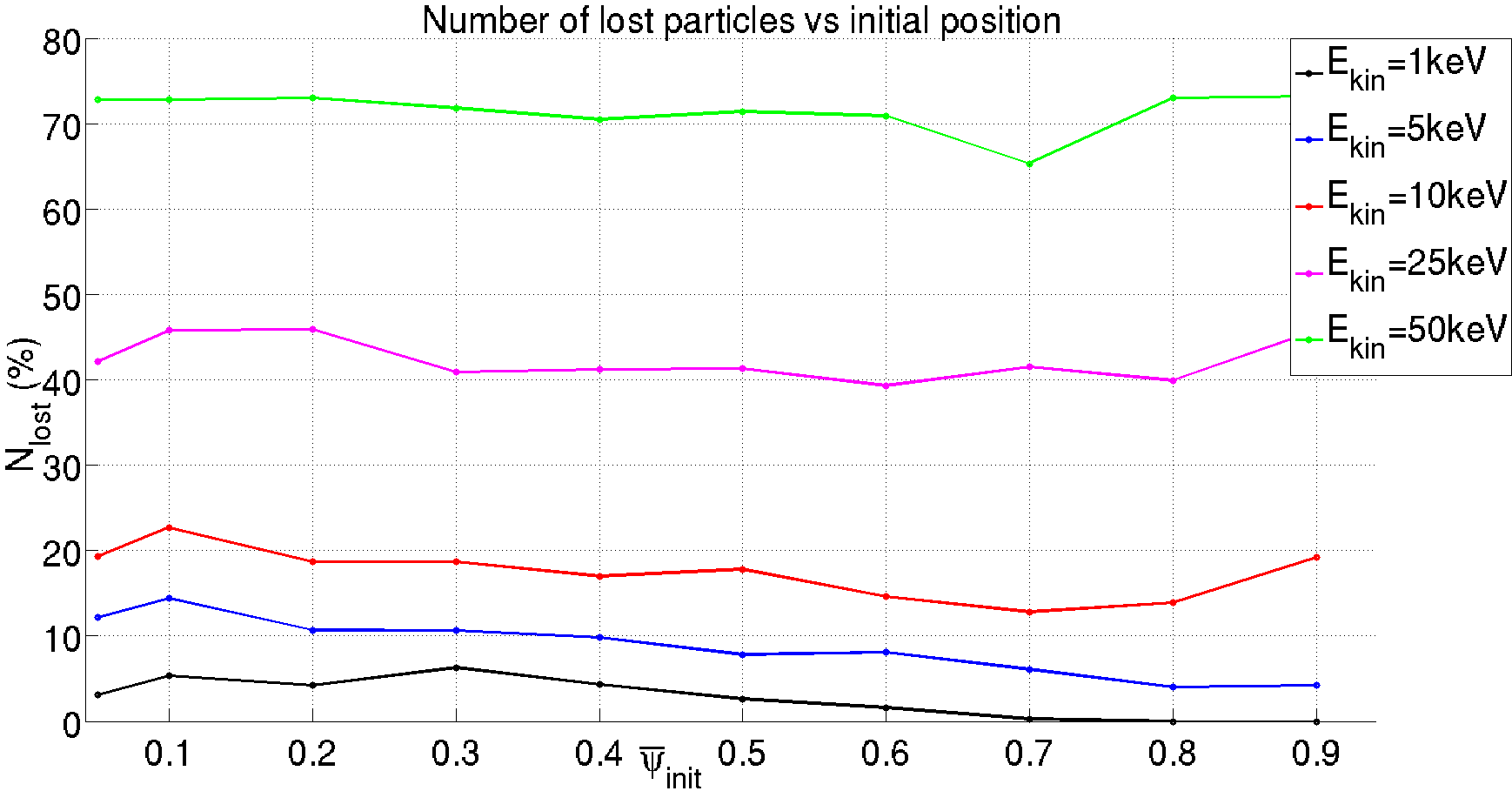} \label{fig:chp4_RE_loss_psiinitvsEkininit}}
	\caption{Fraction of electrons having final kinetic energy above 1MeV (top plot) and lost to the PFCs (bottom plot). Black, blue, red, magenta and green lines are respectively associated to initial kinetic energies of $\mathrm{[1,5,10,25,50]keV}$.}
	\label{fig:chp4_REgeneration_and_loss_psiinitvsEkininit}
\end{figure}
Figure \ref{fig:chp4_REgeneration_and_loss_psiinitvsEkininit} upper and lower plots present respectively the fraction of electrons having final kinetic energy above 1MeV and lost to the wall as a function of the initial radial position, for a range of initial kinetic energies. A first observation of the Figure \ref{fig:chp4_REgeneration_and_loss_psiinitvsEkininit} upper plot reveals that, independently from the initial kinetic energy, a few $\%$ of each initial electron population reaches runaway conditions. In particular, about $\mathrm{\sim 1 \%}$ of the initial thermal (1keV) population (we recall that the pre-TQ central electron temperature is 1keV) runs away. This is one to two orders of magnitude larger than the RE density needed for carrying the whole plasma current. Since no signs of RE were observed during the particular experiment modelled here, it seems that our model strongly over-estimates the runaway seed production. Different possible reasons for this over-prediction will be studied in the next section. 

A second remark on Figure \ref{fig:chp4_REgeneration_and_loss_psiinitvsEkininit} is on the augmentation of the runaway and lost electron fractions with $\mathrm{E_{kin}}$. While the first one is related to the decrease of collision drag at higher kinetic energies, the second is probably linked to the faster particle transport shown in \cite{sommariva18}. However, in these simulations particle losses are lower than the ones given in \cite{sommariva18} for each initial energy level, e.g., less than $10\%$ of the $\mathrm{\left\{\bar{\psi}_{init}=0.05,E_{kin}=10keV \right\}}$ distribution is lost in Figure \ref{fig:chp4_REgeneration_and_loss_psiinitvsEkininit} against the $50\%$ reported in \cite{sommariva18} (Figure 7 upper plot). These discrepancies are probably related to the presence of regions having decelerating $\mathrm{E_{\parallel,eff}}$ which negatively afflict the particle transport. Finally, it has to be recognised that for $\mathrm{E_{kin}<10keV}$ the number of REs reduces considerably when the initial radius increases, supporting the idea that the plasma core has the most favourable conditions for the generation of runaways, essentially because the collision drag is smaller in the core due to the smaller density. This spatial dependency considerably reduces for initial energies above 10keV due to the electron smaller sensitivity to the collisions.
 \section{Possible factors causing a runaway electron over-generation in the JOREK simulated disruptions} \label{chp4_plasma_resistivity_and_RE}

%Four different hypotheses to explain the RE over-generation observed in Section \ref{chp4_electron_acceleration_sec_electron_accelerarion_TQ} are discussed hereafter. The first is related to the particle late initialisation in the disruption simulation. As reported in Table \ref{tab:chp4_Eff_phi45E1to100keVpitch170_tearingtoTQ} second row and in \cite{sommariva18}, the magnetic field stochasticity and the MHD activity start well before the TQ phase probably enhancing particle transport and losses. The second (Subsection \ref{chp4_sub1_high_Z_pollutants}) concerns the absence of high-Z impurities, such as tungsten (W), in the treated disruption simulation. Indeed, high-Z pollutants coming from the metallic walls have the effect to increase the collision drag possibly reducing the electron acceleration. The third hypothesis is related to the slower density rise observed in JOREK simulations than in experiments which would lead to an underestimated drag. Indeed, \cite{nardon17} report that the MGI-induced density increase obtained using the JOREK code is approximately two times smaller than the measured one. This possibility is the subject of Subsection \ref{chp4_sub2_slow_density_increase}. Finally, an evaluation of the influence of the plasma resistivity setting used in the MHD disruption simulation on the RE generation is presented in Subsection \ref{chp4_sub3_plasma_resistivity_and_RE}. This study is justified by the higher plasma resistivity used in JOREK simulations than the JET estimated one which may increase the driving electric field thus, the production of REs.
% 
Three different hypotheses to explain the RE over-generation observed in Section \ref{chp4_electron_acceleration_sec_electron_accelerarion_TQ} are discussed in this Section. The first one (Subsection \ref{chp4_sub1_high_Z_pollutants}) concerns the absence of high-Z impurities, such as tungsten (W), in the treated disruption simulation. Indeed, high-Z pollutants coming from the metallic walls have the effect to increase the collision drag possibly reducing the electron acceleration. The second hypothesis is related to the slower density rise observed in JOREK simulations than in experiments which would lead to an underestimated drag. Indeed, \cite{nardon17} reports that the MGI-induced density increase obtained using the JOREK code is approximately two times smaller than the measured one. This possibility is the subject of Subsection \ref{chp4_sub2_slow_density_increase}. Finally, an evaluation of the influence of the plasma resistivity setting used in the MHD disruption simulation on the RE generation is presented in Subsection \ref{chp4_sub3_plasma_resistivity_and_RE}. This study is justified by the higher plasma resistivity used in JOREK simulations than the JET estimated one which may increase the driving electric field thus, the production of REs.
\subsection{Collision drag due to high-Z pollutants into the plasma discharge}\label{chp4_sub1_high_Z_pollutants}

As introduced above, the presence of high-Z pollutant in the discharge may reduce or avoid the run-away process. Indeed, if W impurities pollute a background plasma, fast electrons have a non-zero probability to collide with them, resulting in an increase of the test particle drag force (Eq.\ref{eq_chp4_drag_force_fussmann} of Section \ref{chp4_sec_collisional_drag}). In order to take into account electron-W interactions, the following terms have to be added respectively to Eq.\ref{chp4_eq_electron_collisional_density} and Eq.\ref{chp4_eq_ion_collisional_density} of Section \ref{chp4_sec_collisional_drag} \cite{solis15}:

\numparts
\begin{gather}
\mathrm{\alpha_{eb,W} = n_{eb,W}{\;}\ln\left(\Lambda_{eb,W}\right)} \\
\mathrm{\alpha_{i,W} = n_W\left[{<Z_W>}^2{\;}\ln\left(\Lambda_{<Z>,W}\right) +Z_{nucl,W}^2{\;}\ln\left(\Lambda_{Z_{nucl,W}}\right) \right]} \\
\mathrm{\Lambda_{eb,W} = \left(\gamma-1\right)\sqrt{\gamma+1}\frac{E_0}{I_W}} \\
\mathrm{\Lambda_{<Z>} = \frac{\lambda_D}{{<Z>}r_e}\frac{I_W}{E_0}}, \quad \mathrm{\Lambda_{Z_{nucl,W}} = \frac{\gamma^2-1}{\gamma}\frac{E_0}{I_W}} 
\end{gather}
\endnumparts  

where $\mathrm{n_{eb,W}}$ and $\mathrm{n_{W}}$ are respectively the W bounded electron and ion number densities, $\mathrm{<Z_W>}$ is the average W charge state, $\mathrm{Z_{nucl,W}=74}$ is the W nuclear charge \cite{linstrom17}, $\mathrm{I_W = 7.86403}$eV is the W ground ionisation energy \cite{linstrom17}, $\mathrm{\lambda_D}$, $\mathrm{r_e}$ and $\mathrm{E_0}$ are respectively the plasma Debye length, the electron classical radius and rest energy as defined in Section \ref{chp4_sec_collisional_drag}. Using this model, the collision drag experienced by an electron having kinetic energy ($\mathrm{E_{kin}}$) varying from 1keV to 10keV and a pitch angle of ${\theta=170^{\circ}}$ on a constant and uniform W distribution ($\mathrm{n_W = 10^{16} (m^{-3})}$) is calculated using Eq.\ref{chp4_eq_collision_drag} of Section \ref{chp4_sec_electric_field_disruption} (where $\mathrm{\alpha_e}$ and $\mathrm{\alpha_i}$ are substituted with $\mathrm{\alpha_{eb,W}}$ and $\mathrm{\alpha_{i,W}}$) and then used to evaluate the W-induced $\mathrm{E_{\parallel,eff}}$ ($\mathrm{E_{\parallel,eff,W}}$). The plasma parameters used for calculating $\mathrm{\lambda_D}$ are the pre-TQ ones ($\mathrm{n=3 \cdot 10^{19} (m^{-3})}$ and $\mathrm{T=1.26keV}$) for the TQ phase and $\mathrm{n=6 \cdot 10^{19} (m^{-3})}$ and $\mathrm{T=270eV}$ for the CQ \cite{sommariva17}. The bounded electron number density is estimated as $\mathrm{n_{eb}=n_w(Z_{nucl,W}-<Z>)}$. A scan in $\mathrm{<Z>}$ from neutral ($\mathrm{<Z>=0}$) to fully ionised ($\mathrm{<Z>=74}$) impurities shows that the largest W contribution to the $\mathrm{E_{\parallel,eff}}$ is of: $\mathrm{E_{\parallel,eff,W}=9.6 (V/m)}$ for the $\mathrm{E_{kin}=1keV}$ case and of $\mathrm{E_{\parallel,eff,W}=1.11 (V/m)}$ for 10keV electrons. A comparison of these estimates with the $\mathrm{E_{\parallel,eff}}$ given in the Table \ref{tab:chp4_Eff_phi45E1to100keVpitch170_tearingtoTQ} of Section \ref{chp4_sec_electric_field_disruption} shows that a possible W pollution of the background plasma has the double effect of decreasing the TQ accelerating $\mathrm{E_{\parallel,eff}}$ (Table \ref{tab:chp4_Eff_phi45E1to100keVpitch170_tearingtoTQ} third row) and of increasing the RE energy threshold at the CQ beginning (Table \ref{tab:chp4_Eff_phi45E1to100keVpitch170_tearingtoTQ} last row). While the latter can be reduced by a factor of two, the TQ driving $\mathrm{E_{\parallel,eff}}$ is orders of magnitude higher than $\mathrm{E_{\parallel,eff,W}}$ thus, the absence of a W impurity background in our simulation can reduce but not completely suppress the RE over-generation observed in Figure \ref{fig:chp4_REgeneration_and_loss_psiinitvsEkininit} upper plot. This conclusion is also supported by the estimation of the W density required to completely suppress the CQ RE production which, for an electron kinetic energy of 60keV (from Figure \ref{fig:chp4_EkinPparProf_psi0e05_Ekin1keV_pitch170} upper plot) is approximately 30 times higher than the plausible $\mathrm{n_W}$ value used above. Anyway, it has to be noted that these considerations are valid only from the electron kinetic point of view. Indeed, modifications of the TQ MHD dynamics, thus to the $\mathrm{E_{\parallel,eff}}$, in disruption simulations caused by the introduction of a high-Z impurity background are difficult to predict and may completely alter the number of REs obtained in simulations. 

\subsection{Estimations of the electron drag force reduction due to slow MGI-induced plasma density increase} \label{chp4_sub2_slow_density_increase}

In this Subsection, we analyse the slower-than-in-experiment plasma density increase as a possible explanation for the RE over-production. This is justified by the evidences reported in \cite{nardon17} (Figure 2) which show that the measured augmentation of the line integrated density (related to the assimilation of the MGI neutral gas) is higher in experiments than in JOREK simulations. For doing so, a scan of the background plasma contribution to the $\mathrm{E_{\parallel,eff}}$ as a function of its density is performed using Eqs.\ref{eq_chp4_drag_force_fussmann}, \ref{chp4_eq_electron_collisional_density} and \ref{chp4_eq_ion_collisional_density} of Section \ref{chp4_sec_collisional_drag} (the TQ and CQ reference plasma temperatures are respectively of 1.26keV and 270eV while the plasma densities are of $\mathrm{3\cdot10^{19} (m^{-3})}$ and of $\mathrm{6\cdot10^{19} (m^{-3})}$, the deuterium density is set to be zero). This scan reveals that a complete suppression of the TQ driving $\mathrm{E_{\parallel,eff}}$ (Table \ref{tab:chp4_Eff_phi45E1to100keVpitch170_tearingtoTQ} third row) for an electron kinetic energy of 1keV and a pitch angle of $\mathrm{170^{\circ}}$ can be attained adding a drag force corresponding to approximately a hundred times higher plasma density than the pre-TQ one (taken at the magnetic axis). On the other hand, the addition of a roughly 7.5 times higher density during the CQ (and with respect to the CQ one) is sufficient to completely suppress the RE generation (this estimate is obtained considering an electron kinetic energy of 60keV which is higher than the maximum $\mathrm{E_{kin}}$ reported in the Figure \ref{fig:chp4_EkinPparProf_psi0e05_Ekin1keV_pitch170} upper plot). While a hundred times increase of the core plasma density is unlikely to happen, it is reasonable to think that an increase of the CQ plasma density would significantly reduce the fraction of the initial electron population running away. 

\subsection{MHD fields and RE generation dependency on the plasma resistivity used in JOREK}\label{chp4_sub3_plasma_resistivity_and_RE}

As concluding subject, we present a first analysis of the influence of the plasma resistivity used in disruption simulation on the magnetic field, $\mathrm{E_{\parallel,eff}}$ and RE generation. This is motivated by the impossibility of obtaining JET disruption simulations with realistic input parameters. Indeed, the plasma resistivity ($\mathrm{\eta}$) used for obtaining the results presented above is believed to be up to an order of magnitude higher than the estimated JET one. We focus our attention on the central plasma resistivity $\eta_0$ (noting that in JOREK the Spitzer-like plasma resistivity model $\eta(T) = \eta_0 (T/T_0)^{-1.5}$ is used, where $T_0$ is the central plasma temperature). This is one of the key parameters ruling the electric field dynamics and the reformation of closed magnetic surfaces after the TQ. For doing this, we analyse two more JOREK disruption simulations in which the resistivity is increased from $\mathrm{\eta_0=3.85\cdot10^{-7}({\Omega}m)}$ to $\mathrm{\eta_0=3.85\cdot10^{-6}({\Omega}m)}$ and $\mathrm{\eta_0=3.85\cdot10^{-5}({\Omega}m)}$, all other parameters being left unchanged. As before, electrons are initialised just before the magnetic field complete stochastisation and followed until closed magnetic surfaces are reformed (total simulation time of $\mathrm{\simeq 1 ms}$ for $\mathrm{\eta_0=3.85\cdot10^{-6}({\Omega}m)}$ and $\mathrm{\simeq 0.28 ms}$ for $\mathrm{\eta_0=3.85\cdot10^{-5}({\Omega}m)}$).
%
%In this Section, we present a first analysis of the influence the disruption simulation input parameters on the magnetic field, $\mathrm{E_{eff}}$ and RE generation. This is motivated by the impossibility of obtaining JET disruption simulations with realistic input parameters. For example, the initial central resistivity ($\mathrm{\eta_0}$) used in the JOREK simulation analysed above is about a factor of five larger than the realistic one. We focus our attention on $\mathrm{\eta_0}$ (noting that in JOREK the Spitzer-like plasma resistivity model $\eta(T) = \eta_0 (T/T_0)^{-1.5}$ is used, where the $\eta_0$ and $T_0$ are respectively the central plasma resistivity and temperature). This is one of the key parameters ruling the electric field dynamics and the reformation of closed magnetic surfaces after the TQ. For doing this, we analyse two more JOREK disruption simulations in which the resistivity is increased from $\mathrm{\eta_0=10^{-7}}$ (in normalised units) to $\mathrm{\eta_0=10^{-6}}$ and $\mathrm{\eta_0=10^{-5}}$, all other parameters being left unchanged. As before, electrons are initialised just before the magnetic field complete stochastisation and followed until closed magnetic surfaces are reformed (total simulation time of $\mathrm{\simeq 1 ms}$ for $\mathrm{\eta_0=10^{-6}}$ and $\mathrm{\simeq 0.28 ms}$ for $\mathrm{\eta_0=10^{-5}}$).  

\begin{table}[h!]
	%	\centering
	\begin{adjustwidth}{-0.85cm}{}
		\begin{tabular}{|c|c|c|c|c|}
			\toprule
			Poincar\'e TQ & $\mathrm{E_{\parallel,eff}=1keV}$ TQ & Poincar\'e CQ & $\mathrm{E_{\parallel,eff}=1keV}$ CQ \\
			\midrule
			\includegraphics[width=3.5cm, height=4.75cm]{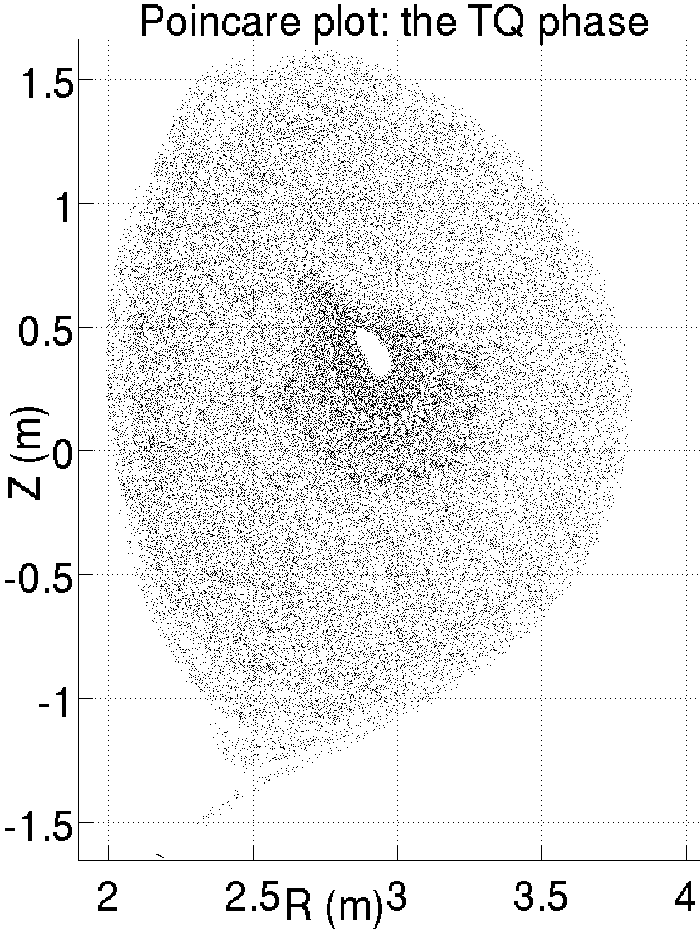} & \includegraphics[width=3.75cm, height=4.75cm]{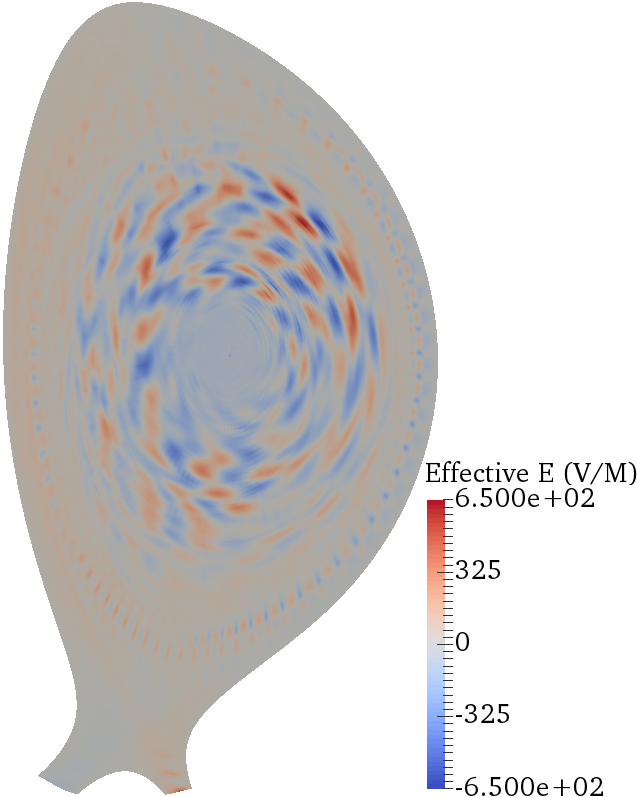} & \includegraphics[width=3.5cm, height=4.75cm]{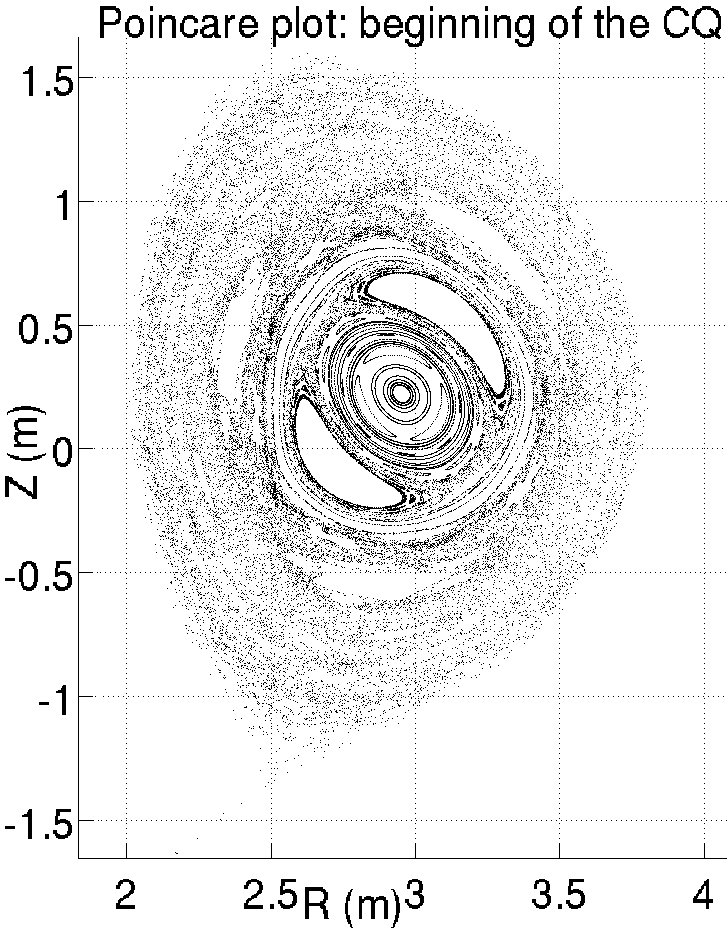} & \includegraphics[width=3.75cm, height=4.75cm]{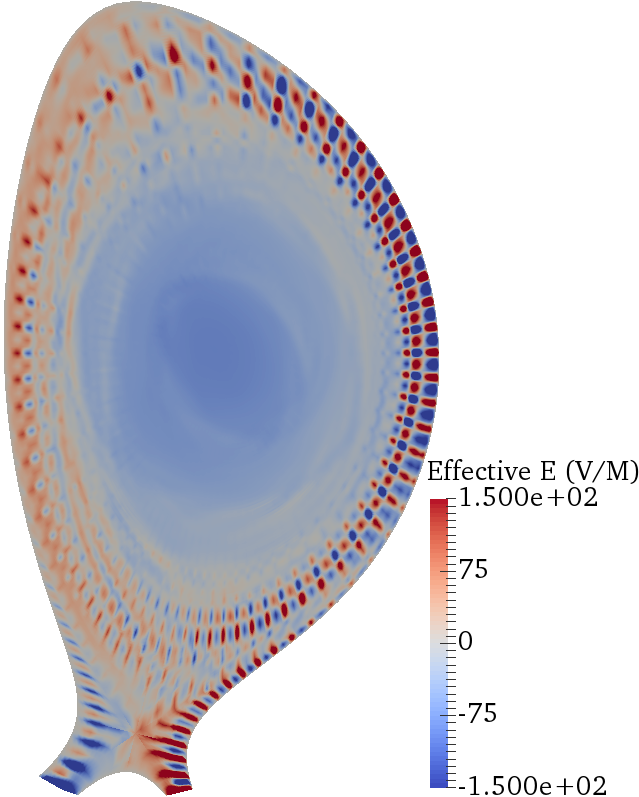} \\
			\midrule
			\includegraphics[width=3.5cm, height=4.75cm]{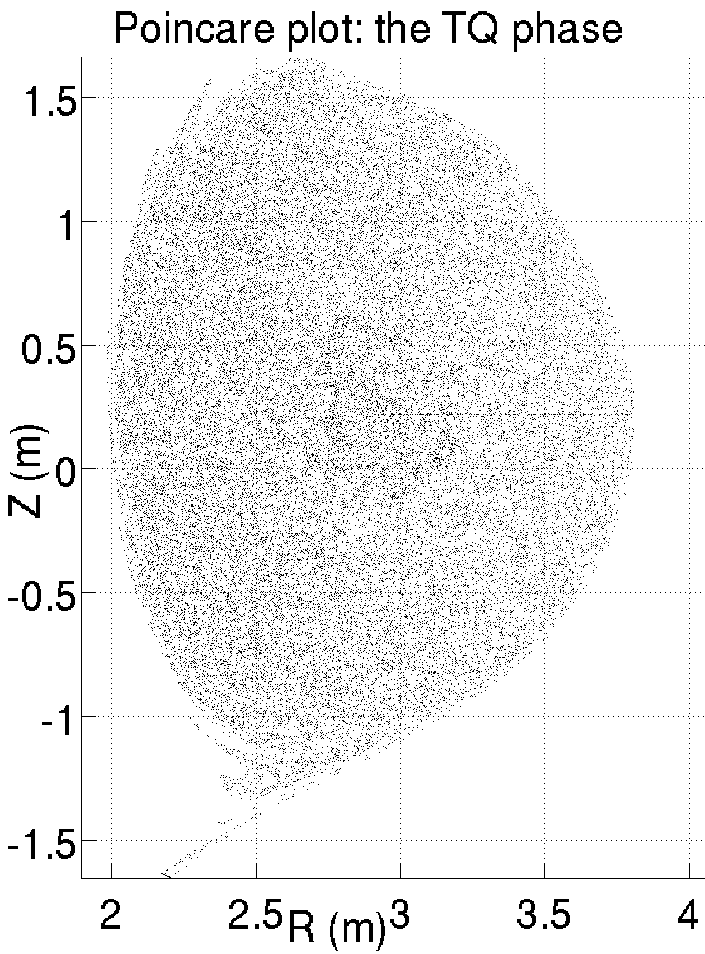} & \includegraphics[width=3.75cm, height=4.75cm]{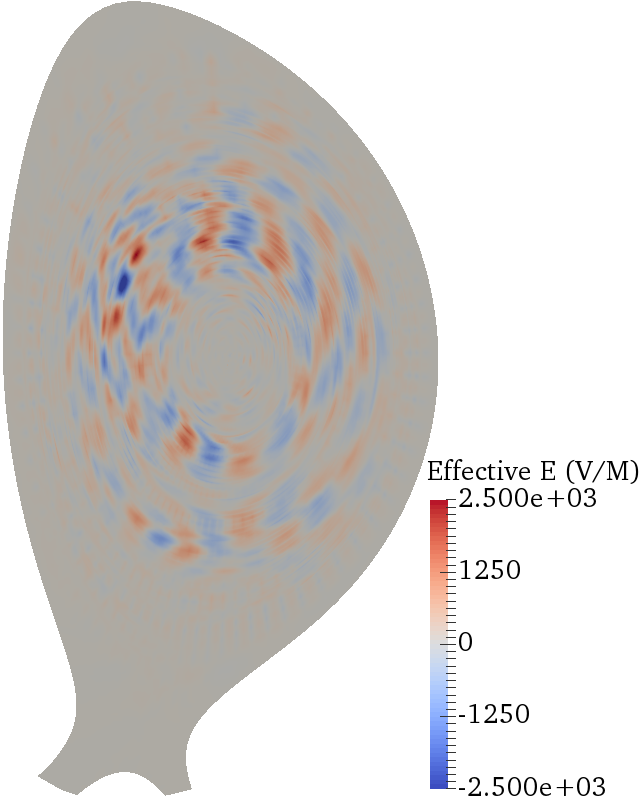} & \includegraphics[width=3.5cm, height=4.75cm]{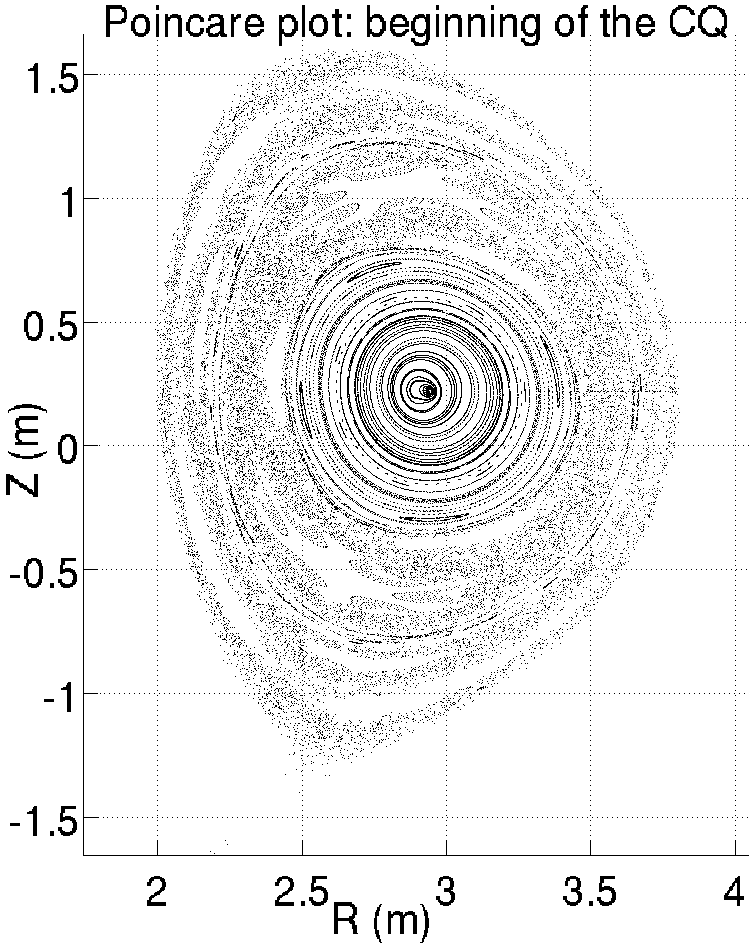} & \includegraphics[width=3.75cm, height=4.75cm]{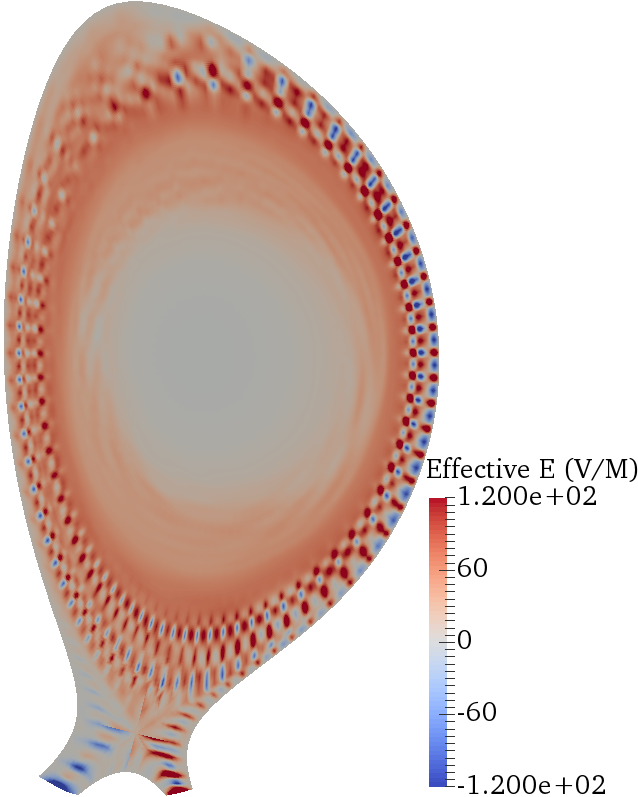} \\
			\midrule
			\includegraphics[width=3.5cm, height=4.75cm]{poincare_nlines300_nurn200_npoint1000_rst4110_chp4.png} & \includegraphics[width=3.75cm, height=4.75cm]{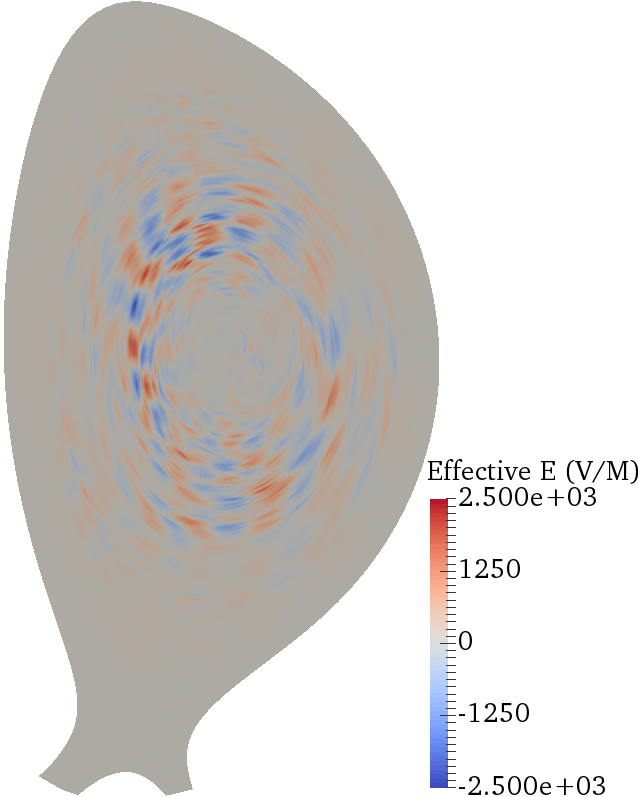} & \includegraphics[width=3.5cm, height=4.75cm]{poincare_nlines300_nurn200_npoint1000_rst8000_chp4.png} & \includegraphics[width=3.75cm, height=4.75cm]{vtk_Epareff_phi45_Ekin1en3_pitch170_rst8000_SI_mod.png} \\
			\bottomrule
		\end{tabular}
	\end{adjustwidth}
	\caption{Poincar\'e and $\mathrm{E_{\parallel,eff}}$ ($\mathrm{\frac{V}{m}}$) ($\mathrm{E_{kin}=1keV,\theta = 170^{\circ}}$) plots of the TQ and CQ phases for simulations with different plasma resistivities: top, middle and bottom plots refer respectively to $\eta_0$ of $\mathrm{3.85\cdot10^{-5}({\Omega}m)}$, $\mathrm{3.85\cdot10^{-6}({\Omega}m)}$ and $\mathrm{3.85\cdot10^{-7}({\Omega}m)}$.}
	\label{tab:chp4_Eff_phi45E1keVpitch170_resistivity_TQCQ}
\end{table}

\begin{figure}[h!]
	\centering
	\includegraphics[width=11.5cm, height=6.25cm]{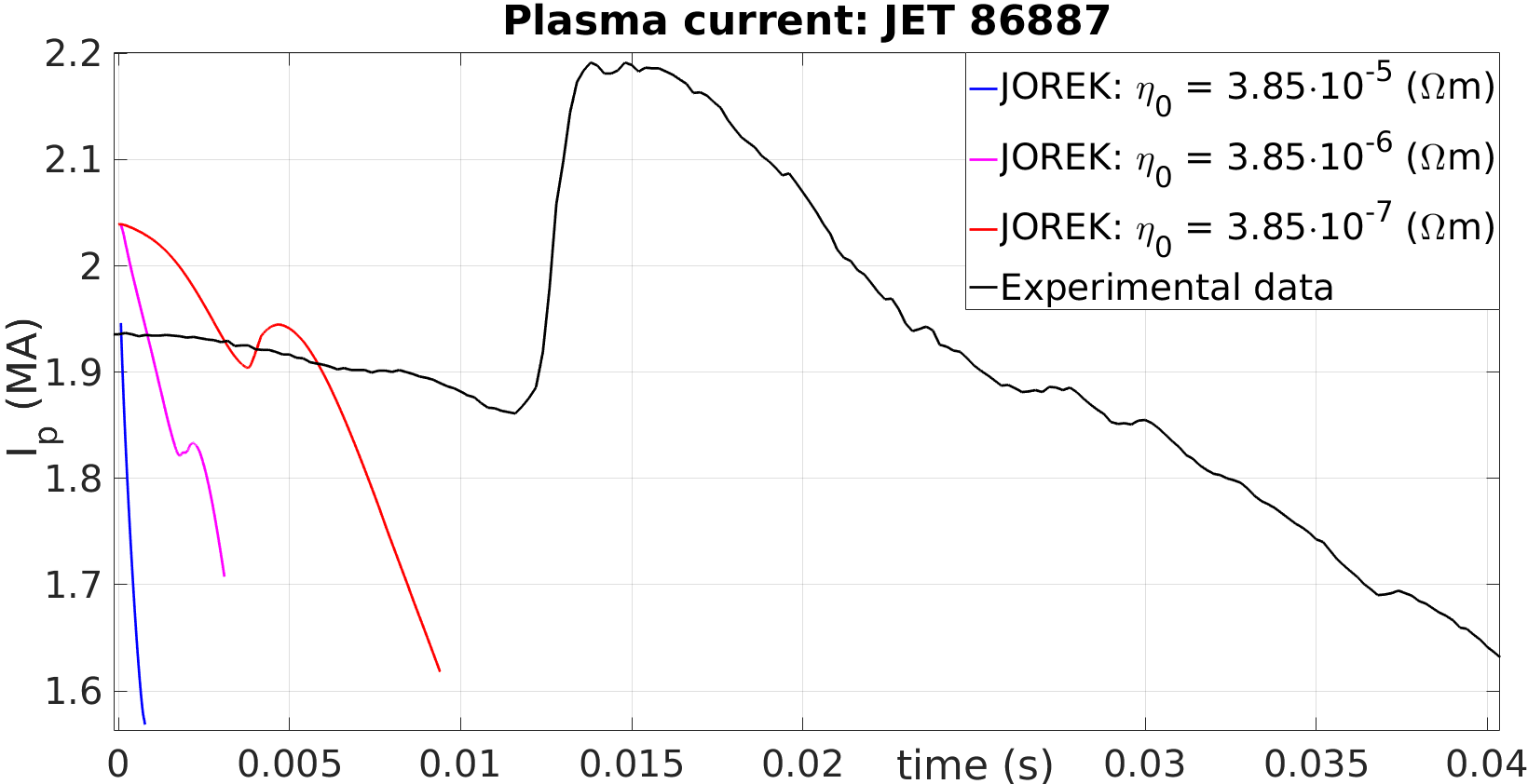}
	\caption{Plasma current from the first gas-plasma interactions to the CQ beginning: the black line represents the experimental data while blue, magenta and red lines denote respectively the $\mathrm{3.85\cdot10^{-5}({\Omega}m)}$, $\mathrm{3.85\cdot10^{-6}({\Omega}m)}$ and $\mathrm{3.85\cdot10^{-7}({\Omega}m)}$ JOREK MHD simulations.}
	\label{fig:chp4_Ip_multiResitivity}
\end{figure}

In Table \ref{tab:chp4_Eff_phi45E1keVpitch170_resistivity_TQCQ} the Poincar\'e and $\mathrm{E_{\parallel,eff}}$ plots for the JOREK simulations obtained using plasma resistivities of $\mathrm{3.85\cdot10^{-5}({\Omega}m)}$, $\mathrm{3.85\cdot10^{-6}({\Omega}m)}$ and $\mathrm{3.85\cdot10^{-7}({\Omega}m)}$ (respectively upper, middle and bottom rows) during the disruption TQ (first and second columns) and CQ (third and fourth columns) phases are reported. The TQ Poincar\'e plots (fist column) show that the magnetic field is globally chaotic independently from the resistivity. Similarities are found in $\mathrm{E_{\parallel,eff}}$ (second column of Table \ref{tab:chp4_Eff_phi45E1keVpitch170_resistivity_TQCQ}). Indeed, in all three cases $\mathrm{E_{\parallel,eff}}$ has a cellular topology composed of poloidally alternating accelerating and decelerating cells. However, while the $\mathrm{E_{\parallel,eff}}$ magnitude is similar for the $\mathrm{\eta_0=3.85\cdot10^{-6}({\Omega}m)}$ and $\mathrm{\eta_0=3.85\cdot10^{-7}({\Omega}m)}$ cases, it is much smaller for $\mathrm{\eta_0=3.85\cdot10^{-5}({\Omega}m)}$. One can also note the presence of an accelerating electric field at the plasma core for the $\mathrm{\eta_0=3.85\cdot10^{-5}({\Omega}m)}$ case which is not present in the other cases. These results show that disruption simulations obtained using the JOREK code recover the insensitivity of the TQ MHD activity with respect to the plasma resistivity already observed in \cite{diamond84} and references therein, at least for the reasonable values of $\mathrm{\eta_0=3.85\cdot10^{-6}({\Omega}m)}$ and of $\mathrm{\eta_0=3.85\cdot10^{-7}({\Omega}m)}$. The magnetic and $\mathrm{E_{\parallel,eff}}$ topologies during the CQ, respectively third and fourth columns of Table \ref{tab:chp4_Eff_phi45E1keVpitch170_resistivity_TQCQ}, vary significantly with $\eta_0$. Indeed, while all simulations display the reformation of closed magnetic surfaces at the center, differences are visible in the Poincar\'e cross sections and, more importantly, on $\mathrm{E_{\parallel,eff}}$. The latter transits from a strong accelerating to a decelerating configuration with the decrease of plasma resistivity. This changeover is explained by the slower $\mathrm{I_p}$ decay induced by the lower resistivity, which is visible in Figure \ref{fig:chp4_Ip_multiResitivity} displaying the experimental and JOREK simulated $\mathrm{I_p}$ traces. Summarising, Table \ref{tab:chp4_Eff_phi45E1keVpitch170_resistivity_TQCQ} shows that, in this case of study, the CQ $\mathrm{E_{\parallel,eff}}$ is strongly sensitive to the choice of $\eta_0$. Conversely, initial resistivity variations weakly influence the CQ magnetic configuration and the TQ fields, especially if we compare the $\mathrm{\eta_0=3.85\cdot10^{-6}({\Omega}m)}$ and $\mathrm{\eta_0=3.85\cdot10^{-7}({\Omega}m)}$ cases.    

\begin{figure}[h!]
		\centering
		\includegraphics[width=11cm, height=6.25cm]{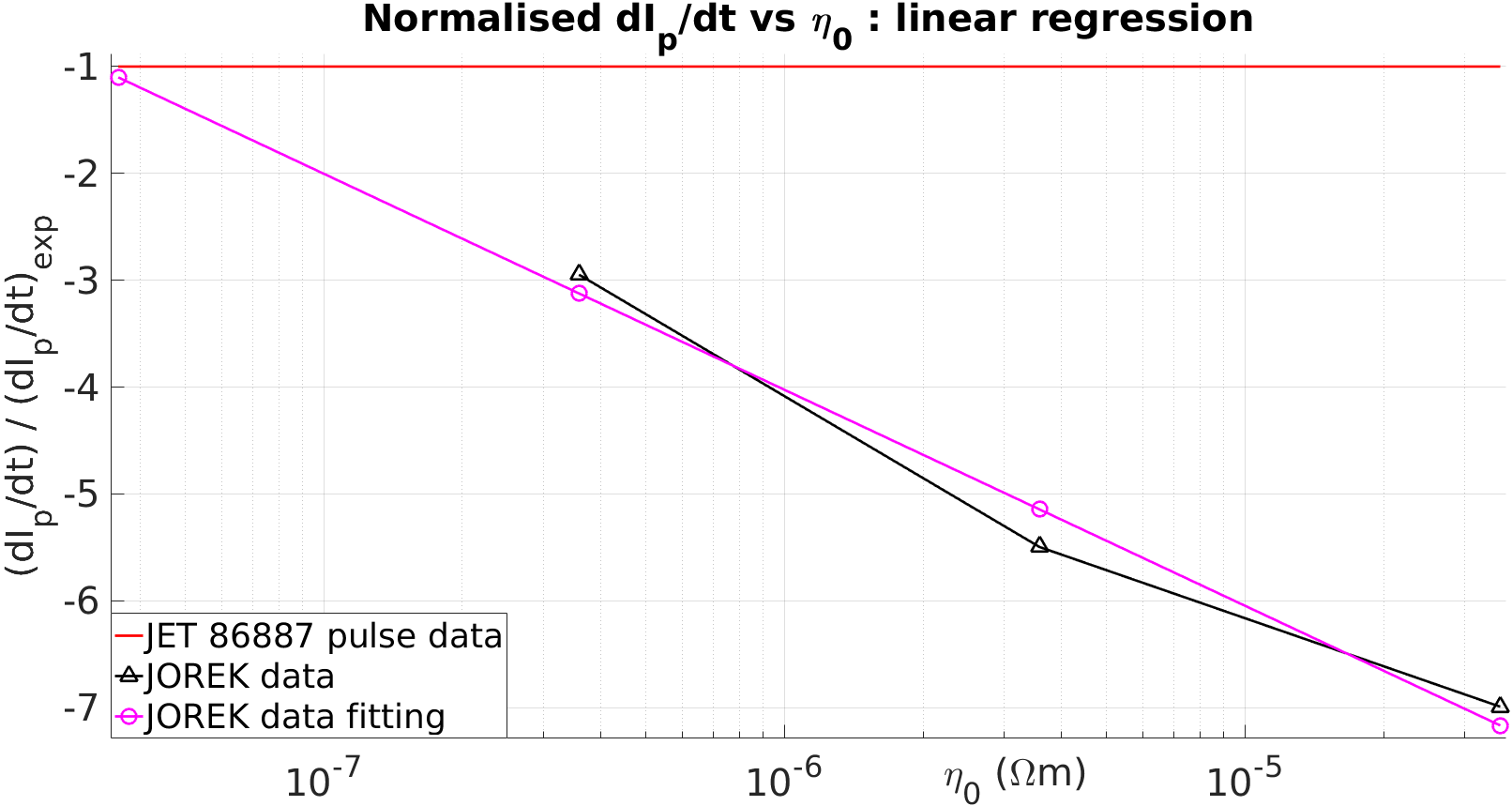}
		\caption{CQ $\mathrm{\frac{d I_p}{d t}}$ linear regression (normalised to the experimental value) as a function of the plasma resistivity: red, black and magenta lines represent respectively the experimental value, JOREK simulations and their extrapolation to the experimental $\eta_0$.}
		\label{fig:chp4_LinearRegression_dIpdt_multiResitivity}
\end{figure}

Analysing Figure \ref{fig:chp4_Ip_multiResitivity}, one can see that the experimental $\mathrm{I_p}$ decay is slower than the simulated one, as could be expected from the artificial increase of resistivity in the simulations. This probably partly explains the RE overproduction observed in Section \ref{chp4_electron_acceleration_sec_electron_accelerarion_TQ}. Thus, it is interesting to try to extrapolate the JOREK results to the realistic level of $\mathrm{\eta_0 = 7.7 \cdot 10^{-8}({\Omega}m)}$. For this purpose, we estimated the current decay rate via linear regression of the CQ $\mathrm{I_p}$ profiles exhibited in Figure \ref{fig:chp4_Ip_multiResitivity} and, then, we extrapolated them to $\mathrm{\eta_0=7.7 \cdot 10^{-8}({\Omega}m)}$ via logarithmic fitting. The result is reported in Figure \ref{fig:chp4_LinearRegression_dIpdt_multiResitivity} which suggests that the experimental $\mathrm{I_p}$ decay rate would be recovered if the realistic resistivity could be used.

As a side remark on Figure \ref{fig:chp4_LinearRegression_dIpdt_multiResitivity}, the JOREK $\mathrm{\frac{d I_p}{d t}}$ (black line) is not directly proportional to $\eta_0$. We presume that this is due to the CQ plasma temperature increasing with $\eta_0$ due to a larger Ohmic heating.

\begin{figure}[h!]
	\centering
	\includegraphics[width=11.5cm, height=6.25cm]{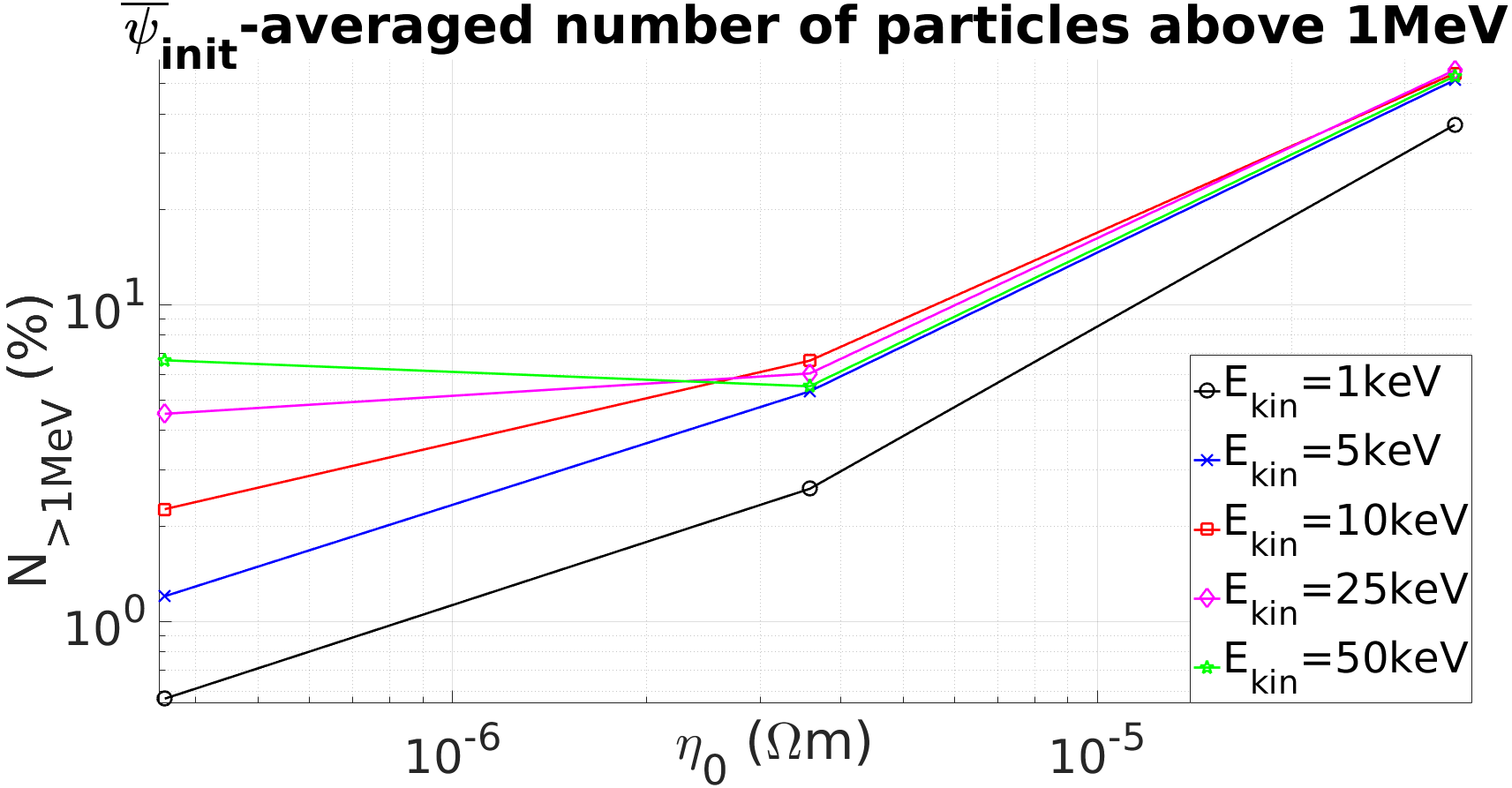}
	\caption{Fraction of electrons having final kinetic energy above 1MeV as a function of the initial plasma resistivity for different initial kinetic energies (averaged over all initial radial positions). Black, blue, red, magenta and green lines correspond respectively to initial $\mathrm{E_{kin}}$ of $\mathrm{1keV}$, $\mathrm{5keV}$, $\mathrm{10keV}$, $\mathrm{25keV}$ and $\mathrm{50keV}$}
	\label{fig:chp4_REgeneration_multieta}
\end{figure}

The fraction of electrons becoming RE (averaged over the initial positions) as a function of $\mathrm{\eta_0}$ for different initial $\mathrm{E_{kin}}$ is given in Figure \ref{fig:chp4_REgeneration_multieta}. This figure shows that a decrease of $\mathrm{\eta_0}$ reduces the number of produced RE. In particular, the fraction of RE generated from the 1keV populations drops from $\mathrm{\sim 37\%}$ for $\mathrm{\eta_0=3.85\cdot10^{-5}({\Omega}m)}$ to $\mathrm{\sim 2.6\%}$ and $\mathrm{\sim 0.6\%}$ respectively for $\mathrm{\eta_0=3.85\cdot10^{-6}({\Omega}m)}$ and $\mathrm{\eta_0=3.85\cdot10^{-7}({\Omega}m)}$. However, it seems that for $\mathrm{\eta_0=7.7 \cdot 10^{-8}({\Omega}m)}$ the fraction of RE would still be significant compared to the $10^{-1} - 10^{-2}\%$ required for carrying the whole plasma current. It is worth remarking that the dependence between the RE number and the initial plasma resistivity for the range $\mathrm{\eta_0 \in [3.85\cdot10^{-7},3.85\cdot10^{-6}]({\Omega}m)}$ weakens when the $\mathrm{E_{kin}}$ is increased.      

\section{Summary and conclusions} \label{chp4_summary}

In order to study the generation of fast electrons, a drag force, modeling collisions between relativistic electrons and a background plasma containing $\mathrm{D_2}$ neutrals, is introduced in the GC pusher of the JOREK fast particle tracker.

After having introduced the parallel effective electric field (electric force plus drag force), we analysed its evolution during the treated disruption simulation for kinetic energies of $\mathrm{E_{kin}=[1,10,100]keV}$. Before (t=3.55ms) and after (t=6.94ms) the TQ, the 1keV $\mathrm{E_{\parallel,eff}}$ is dominated by the drag force whereas at higher energies a transition towards an $\mathrm{E_{\parallel}}$ dominated $\mathrm{E_{\parallel,eff}}$ is observed. In contrast, throughout the TQ, $\mathrm{E_{\parallel}}$ dominates the drag force independently from the initial $\mathrm{E_{kin}}$. During this phase, the $\mathrm{E_{\parallel,eff}}$ fluctuations reach intensities up to $\sim$2kV/m and have a topology characterised by poloidally alternated accelerating and decelerating cells. Moreover, these cells shrink in size and extend from the plasma core to the edge with time. The origin of this field is related to the strong MHD activity taking place during the TQ but, at the moment, the precise mechanisms remain to be investigated.

Then, we used test particles simulations in order to analyse the generation of fast electrons. Results show that the $\mathrm{E_{\parallel}}$ activity taking place during the TQ causes an important spreading of the momentum space particle distribution in counter and co-plasma current directions. Considering the counter-$\mathrm{I_p}$ accelerated particles, a few $\%$ of them reach kinetic energies at which $\mathrm{E_{\parallel,eff}}$ remains dominated by the electric field after the TQ while remaining within the plasma core region. After the TQ, these electrons are confined by the reformation of closed magnetic surfaces and driven to RE energies during the CQ by the inductive electric field. The fate of the non-RE electrons strongly depends on the population initial energy, i.e., particles having high initial $\mathrm{E_{kin}}$ are generally lost to the wall while at low $\mathrm{E_{kin}}$ electron thermalisation is the dominant process. 

In the JET 86887 disruption experiment, no RE were observed. In contrast, the JOREK particle simulation indicates a strong generation of RE even for initially thermal electron populations. Three different possible reasons explaining this discrepancy have been addressed: the absence of high-Z (tungsten) impurities in the simulation, a plasma density increase slower in simulations than in experiments and the fact that the JOREK disruption simulations are run with a plasma resistivity significantly higher than the JET estimated one. While the tungsten concentration required to completely suppress the CQ RE production is found to be unrealistically elevated, a 7.5 times higher plasma density should be sufficient to avoid the run-away process. This last result has to be considered in concert with the scan in $\mathrm{\eta_0}$ conducted in order to asses its importance on both MHD and particle dynamics. Indeed, simulations show that, while the TQ $\mathrm{E_{\parallel,eff}}$ is weakly affected by $\mathrm{\eta_0}$ variations (in agreement with the discussions reported in \cite{diamond84} and references therein), the CQ $\mathrm{E_{\parallel,eff}}$ (at $\mathrm{E_{kin}=1keV}$) varies significantly with $\mathrm{\eta_0}$, i.e., for the very high resistivity of $\mathrm{\eta_0=3.85\cdot10^{-5}({\Omega}m)}$ a strong accelerating $\mathrm{E_{\parallel}}$ is the dominant contribution to $\mathrm{E_{\parallel,eff}}$ whereas for $\mathrm{\eta_0=3.85\cdot10^{-7}({\Omega}m)}$ the collision drag dominates. An extrapolation of the JOREK results towards a realistic resistivity suggests that the experimental CQ $\mathrm{\frac{dI_p}{dt}}$, thus the $\mathrm{E_{\parallel}}$, would be recovered. Therefore, it is reasonable to hypothesise that the combined effect of both higher plasma density and lower $\mathrm{E_{\parallel}}$ would prevent the generation of REs and, in particular, their acceleration during the CQ. Anyway, it has to be mentioned that the analysed disruption simulation probably presents a weaker MHD activity than in experiment as can be deduced by the small $\mathrm{I_p}$ spike associated to the simulated TQ (Figure \ref{fig:chp4_Ip_multiResitivity}). The consequences of this are not clear and, as a consequence, further efforts are needed to reconcile simulations and experiments.

Despite the quantitative mismatch between simulations and experiments, the present work suggests that a kind of Dreicer generation might take place during the TQ and at the CQ beginning. Indeed, electrons can be accelerated by large parallel electric field associated to the TQ MHD activity and, after a prompt reconfinement due to the reformation of magnetic surfaces, become RE thanks to the subsequent acceleration induced by the CQ inductive electric field. This mechanism, which was not reported before at the best knowledge of the authors, may strongly influence the primary RE seed estimates performed for ITER. Thus, further studies are advisable in order to understand and characterise the nature of the actual TQ MHD activity of a mitigated disruption and its capabilities to generate supra-thermal electron populations.

Considering future developments, the present work suggests a multiplicity of research axes. More advanced theoretical investigations should be performed (possibly with the help of simplified numerical models) in order to better understand the physics underlying the parallel electric field dynamics and electron deconfinement during the TQ. Further numerical experiments using the JOREK code will be also performed for improving the quantitative match with the experiments, in particular, with this JET MGI case. For example, it has been found recently that simulations including an impurity background display larger plasma current spike than the one presented in this paper. Another axis of research consists in testing the ability of codes such as JOREK to qualitatively reproduce robust experimental trends like the RE existence domain in JET as a function of the toroidal magnetic field or the quantity of injected impurities \cite{reux15}, the dependencies on the magnetic configuration (divertor vs limiter) and on the type of MGI gas. In addition, test electron studies will be also repeated for the JOREK shattered pellet injection simulations \cite{hu2017} in order to assess differences and similarities with respect to the MGI cases.

%As reported above, advancements will necessarily require the achievement of disruption simulations being closer to the experimental results. In this respect a new high-Z impurity model is implemented within the JOREK code and it starts furnishing promising results in spite of the difficulties already founded when realistic JET plasma parameters are used. We leave the analysis of the fast electron transport and acceleration in these new disruption simulations for future publications. Another interesting development would be the study of the RE dynamics in simulations where the disruption is caused by shattered pellet injection. Indeed, \cite{hu2017} shows that the shattered pellet injection induced TQ has a far different MHD activity than the MGI one with consequent different electron acceleration and confinement properties.
%
%The validation of the JOREK disruption-RE simulations against JET experimental data is also left as future work. Indeed, a qualitative validation of the numerical results against experiments can be sought even if a quantitative one represent a formidable task. As an example, a scan quantifying the RE production as a function of the magnetic field and the quantity of injected impurities will be performed in the near future and compared with RE existence domain reported in \cite{reux15}. This will also be repeated for the shattered pellet injection case as soon as the related experiments will be performed.
%
Finally, the JOREK fast particle tracker could be improved by implementing a Monte Carlo operator for simulating the electron Coulomb collisions such as the one reported in \cite{sarkimaki17} or in \cite{delCastilloNegrete18}. It has to be remarked that a Monte Carlo approach to collisions will not only improve the phase space dynamics description for particles having nearly thermal energies \cite{fussman79} but it will also allow the evaluation of the electron transport in physical space due to collision scattering.    

%a possible, so far not considered, RE formation mechanism: Dreicer acceleration due to large local parallel electric fields associated to the TQ MHD activity, combined with a prompt reconfinement after the TQ due to the reformation of magnetic flux surfaces and subsequent acceleration by the CQ inductive electric field.

\section{Acknowledgments} \label{Acknowledgments}

The authors wish to thank Daan van Vugt, Gergely Papp and Allen H. Boozer for the fruitful discussions. In addition, we thank the two anonymous reviewers whose comments and suggestions helped improve and clarify this manuscript.

This work has been carried out within the framework of the EUROfusion Consortium and has received funding from the Euratom research and training programme 2014-2018 under grant agreement No 633053. The views and opinions expressed herein do not necessarily reflect those of the European Commission.

Simulations were performed on Curie and Marconi supercomputers respectively at CEA-TGCC and CINECA (MARCONI-Fusion project). We thank these institutions for the resources granted.

\bibliographystyle{ieeetr}
\bibliography{refs}

%Example TeX coding might be:
%\begin{verbatim}
%\documentclass[12pt]{iopart}
%\usepackage{CJK}
%.
%.
%.
%\begin{document}
%\begin{CJK*}{GBK}{ }

%\title[]{Title of article}
%\author{Author Name (CJK characters)}
%\address{Department, University, City, Country}
%.
%.
%.
%\end{CJK*}
%\end{verbatim}

%To avoid potential problems in handling the CJK characters in submissions, authors should always include a PDF of the full version of their %papers (including all figure files, tables, references etc) with the CJK characters in it.

\end{document}